\documentclass[a4paper,12pt]{article}
\usepackage{graphicx}
\usepackage{graphics}
\usepackage{epsf}
\begin{document}
\newcommand{\be}{\begin{equation}}
\newcommand{\beq}{\begin{equation}}
\newcommand{\eeq}{\end{equation}}
\newcommand{\ee}{\end{equation}}

\newcommand{\beqn}{\begin{eqnarray}}
\newcommand{\eeqn}{\end{eqnarray}}
\newcommand{\bea}{\begin{eqnarray}}
\newcommand{\ena}{\end{eqnarray}}
\newcommand{\ra}{\rightarrow}
\newcommand{\susy}{{{\cal SUSY}$\;$}}
\newcommand{\su}{$ SU(2) \times U(1)\,$}

\newcommand{\gag}{$\gamma \gamma$ }
\newcommand{\gagt}{\gamma \gamma }
\newcommand{\gam}{\gamma \gamma }
\def\W{{\mbox{\boldmath $W$}}}
\def\B{{\mbox{\boldmath $B$}}}
\def\V{{\mbox{\boldmath $V$}}}
\newcommand{\np}{Nucl.\,Phys.\,}
\newcommand{\pl}{Phys.\,Lett.\,}
\newcommand{\pr}{Phys.\,Rev.\,}
\newcommand{\prl}{Phys.\,Rev.\,Lett.\,}
\newcommand{\prep}{Phys.\,Rep.\,}
\newcommand{\zp}{Z.\,Phys.\,}
\newcommand{\sovjnp}{{\em Sov.\ J.\ Nucl.\ Phys.\ }}
\newcommand{\nuclinst}{{\em Nucl.\ Instrum.\ Meth.\ }}
\newcommand{\annp}{{\em Ann.\ Phys.\ }}
\newcommand{\intjmp}{{\em Int.\ J.\ of Mod.\  Phys.\ }}

\newcommand{\eps}{\epsilon}
\newcommand{\mw}{M_{W}}
\newcommand{\mww}{M_{W}^{2}}
\newcommand{\mwmw}{M_{W}^{2}}
\newcommand{\mhmh}{M_{H}^2}
\newcommand{\mz}{M_{Z}}
\newcommand{\mzz}{M_{Z}^{2}}

\newcommand{\cw}{\cos\theta_W}
\newcommand{\sw}{\sin\theta_W}
\newcommand{\tw}{\tan\theta_W}
\def\cww{\cos^2\theta_W}
\def\sww{s^2_W}
\def\tww{\tan^2\theta_W}

\newcommand{\epm}{$e^{+} e^{-}\;$}
\newcommand{\epemt}{$e^{+} e^{-}\;$}
\newcommand{\epem}{e^{+} e^{-}\;}
\newcommand{\ememt}{$e^{-} e^{-}\;$}
\newcommand{\emem}{e^{-} e^{-}\;}

\newcommand{\lra}{\leftrightarrow}
\newcommand{\tr}{{\rm Tr}}
\def\ls1{{\not l}_1}
\newcommand{\cms}{centre-of-mass\hspace*{.1cm}}

\newcommand{\ie}{{\em i.e.}}
\newcommand{\cm}{{{\cal M}}}
\newcommand{\cl}{{{\cal L}}}
\newcommand{\cd}{{{\cal D}}}
\newcommand{\cv}{{{\cal V}}}
\def\slashc{c\kern -.400em {/}}
\def\slashp{p\kern -.400em {/}}
\def\slashL{L\kern -.450em {/}}
\def\slashcl{\cl\kern -.600em {/}}
\def\Ww{{\mbox{\boldmath $W$}}}
\def\B{{\mbox{\boldmath $B$}}}
\def\noi{\noindent}
\def\nn{\noindent}
\def\sm{${\cal{S}} {\cal{M}}\;$}
\def\smn{${\cal{S}} {\cal{M}}$}
\def\nph{${\cal{N}} {\cal{P}}\;$}
\def\sb{$ {\cal{S}}  {\cal{B}}\;$}
\def\ssb{${\cal{S}} {\cal{S}}  {\cal{B}}\;$}
\def\ssbe{{\cal{S}} {\cal{S}}  {\cal{B}}}
\def\cviol{${\cal{C}}\;$}
\def\pviol{${\cal{P}}\;$}
\def\cpviol{${\cal{C}} {\cal{P}}\;$}

\newcommand{\sinsq}{\sin^2\theta}
\newcommand{\cossq}{\cos^2\theta}
\newcommand{\yt}{y_\theta}

\def\sinb{\sin\beta}
\def\cosb{\cos\beta}
\def\sinbb{\sin (2\beta)}
\def\cosbb{\cos (2 \beta)}
\def\tgb{\tan \beta}
\def\tgbt{$\tan \beta\;\;$}
\def\tgbsq{\tan^2 \beta}
\def\sinal{\sin\alpha}
\def\cosal{\cos\alpha}
\def\stop{\tilde{t}}
\def\sto{\tilde{t}_1}
\def\stt{\tilde{t}_2}
\def\stl{\tilde{t}_L}
\def\str{\tilde{t}_R}
\def\msto{m_{\sto}}
\def\mstosq{m_{\sto}^2}
\def\mstt{m_{\stt}}
\def\msttsq{m_{\stt}^2}
\def\mt{m_t}
\def\mtsq{m_t^2}
\def\sint{\sin\theta_{\stop}}
\def\sintt{\sin 2\theta_{\stop}}
\def\cost{\cos\theta_{\stop}}
\def\sintsq{\sin^2\theta_{\stop}}
\def\costsq{\cos^2\theta_{\stop}}
\def\mqtt{\M_{\tilde{Q}_3}^2}
\def\mutt{\M_{\tilde{U}_{3R}}^2}
\def\sbottom{\tilde{b}}
\def\sbo{\tilde{b}_1}
\def\sbt{\tilde{b}_2}
\def\sbl{\tilde{b}_L}
\def\sbr{\tilde{b}_R}
\def\msbo{m_{\sbo}}
\def\msbosq{m_{\sbo}^2}
\def\msbt{m_{\sbt}}
\def\msbtsq{m_{\sbt}^2}
\def\mt{m_t}
\def\mtsq{m_t^2}
\def\selectron{\tilde{e}}
\def\seo{\tilde{e}_1}
\def\set{\tilde{e}_2}
\def\sel{\tilde{e}_L}
\def\ser{\tilde{e}_R}
\def\mseo{m_{\seo}}
\def\mseosq{m_{\seo}^2}
\def\mset{m_{\set}}
\def\msetsq{m_{\set}^2}
\def\msel{m_{\sel}}
\def\mser{m_{\ser}}
\def\me{m_e}
\def\mesq{m_e^2}
\def\snu{\tilde{\nu}}
\def\snue{\tilde{\nu_e}}
\def\set{\tilde{e}_2}
\def\snul{\tilde{\nu}_L}
\def\msnue{m_{\snue}}
\def\msnuesq{m_{\snue}^2}
\def\smuon{\tilde{\mu}}
\def\smul{\tilde{\mu}_L}
\def\smur{\tilde{\mu}_R}
\def\msmul{m_{\smul}}
\def\msmulsq{m_{\smul}^2}
\def\msmur{m_{\smur}}
\def\msmursq{m_{\smur}^2}
\def\stau{\tilde{\tau}}
\def\stauo{\tilde{\tau}_1}
\def\staut{\tilde{\tau}_2}
\def\staul{\tilde{\tau}_L}
\def\staur{\tilde{\tau}_R}
\def\mstauo{m_{\stauo}}
\def\mstauosq{m_{\stauo}^2}
\def\mstaut{m_{\staut}}
\def\mstautsq{m_{\staut}^2}
\def\mtau{m_\tau}
\def\mtausq{m_\tau^2}
\def\gluino{\tilde{g}}
\def\mgluino{m_{\tilde{g}}}
\def\mchi{m_\chi^+}
\def\neuto{\tilde{\chi}_1^0}
\def\mneuto{m_{\tilde{\chi}_1^0}}
\def\neutt{\tilde{\chi}_2^0}
\def\mneutt{m_{\tilde{\chi}_2^0}}
\def\neutth{\tilde{\chi}_3^0}
\def\mneutth{m_{\tilde{\chi}_3^0}}
\def\neutf{\tilde{\chi}_4^0}
\def\mneutf{m_{\tilde{\chi}_4^0}}
\def\chargop{\tilde{\chi}_1^+}
\def\mchargo{m_{\tilde{\chi}_1^+}}
\def\chargtp{\tilde{\chi}_2^+}
\def\mchargt{m_{\tilde{\chi}_2^+}}
\def\chargom{\tilde{\chi}_1^-}
\def\chargtm{\tilde{\chi}_2^-}
\def\bino{\tilde{b}}
\def\wino{\tilde{w}}
\def\photino{\tilde{\gamma}}
\def\zino{tilde{z}}
\def\sdowno{\tilde{d}_1}
\def\sdownt{\tilde{d}_2}
\def\sdownl{\tilde{d}_L}
\def\sdownr{\tilde{d}_R}
\def\supo{\tilde{u}_1}
\def\supt{\tilde{u}_2}
\def\supl{\tilde{u}_L}
\def\supr{\tilde{u}_R}
\def\mh{m_h}
\def\mht{m_h^2}
\def\MH{M_H}
\def\MHt{M_H^2}
\def\MA{M_A}
\def\MAt{M_A^2}
\def\MHp{M_H^+}
\def\MHm{M_H^-}
\def\mb{m_b}
\def\m0{M_0}
\def\mhf{M_{1/2}}
\def\dMb{\Delta m_b}
\def\mbmb{m_b(m_b)}
\def\tb{\tan \beta}
\def\gstar{g_*^{1/2}}
\def\bbar{b\overline{b}}
\def\ttbar{t\overline{t}}
\def\ccbar{c\overline{c}}
\def\micro{{\tt micrOMEGAs}}
\def\darksusy{{\tt DarkSUSY}}
\def\comphep{{\tt CalcHEP}}
\def\isasugra{{\tt ISASUGRA/Isajet}}
\def\isajet{{\tt Isajet}}
\def\suspect{{\tt Suspect}}
\def\softsusy{{\tt SOFTSUSY}}
\def\spheno{{\tt Spheno}}
\def\bsgamma{b\ra s\gamma}
\def\bsmu{B_s\ra \mu^+\mu^-}
\def\micronew{{\tt micrOMEGAs1.3}}
\def\calchep{{\tt CalcHEP}}
\def\feynhiggs{{\tt FeynHiggsFast}}
\def\gmuon{{(g-2)_\mu}}
\def\ma{M_A}
\def\delrho{\Delta\rho}

\newcommand{\msq}{\tilde{m}}
\newcommand{\Pop}[1]{Q_{#1}}
\newcommand{\Gop}[1]{Q_{#1}}
\newcommand{\msd}[1]{m_{\tilde{d}_{#1}}}
\newcommand{\msb}[1]{m_{\tilde{b}_{#1}}}
\newcommand{\mst}[1]{m_{\tilde{t}_{#1}}}
\newcommand{\msu}[1]{m_{\tilde{u}_{#1}}}
\newcommand{\mug}{\mu_{\tilde g}}
\newcommand{\msbar}{\overline{\rm MS}}
\newcommand{\cb}{\cos \beta}
\def \ct  {c_{\tilde{t}}}
\def \st  {s_{\tilde{t}}}
\def\f{\frac}
\def\as{\alpha_s}
\def\al{\alpha_s}
\def\mg{m_{\tilde{g}}}
\baselineskip=18pt

\bibliographystyle{unsrt}
\begin{titlepage}
\def\baselinestretch{1.2}
\topmargin     -0.25in

\vspace*{\fill}
\begin{center}
{\large {\bf micr\Large{OMEGA}s: Version 1.3
 }} \vspace*{0.5cm}

\begin{tabular}[t]{c}

{\bf G.~B\'elanger$^{1}$, F.~Boudjema$^{1}$,  A. Pukhov$^{2}$,
A. Semenov$^{3}$}
 \\
\\
\\
{\it 1. Laboratoire de Physique Th\'eorique}
{\large LAPTH}
\footnote{URA 14-36 du CNRS, associ\'ee  \`a
l'Universit\'e de Savoie.}\\
 {\it Chemin de Bellevue, B.P. 110, F-74941 Annecy-le-Vieux,
Cedex, France.}\\

{\it 2. Skobeltsyn Institute of Nuclear Physics,
Moscow State University} \\ {\it Moscow 119992,
Russia }\\

{\it 3. Joint Institute for Nuclear Research (JINR)}\\
{\it 141980, Dubna, Moscow Region, Russia}\\

\end{tabular}
\end{center}

\centerline{ {\bf Abstract} }
\baselineskip=14pt
\noindent
 {\small We present the latest version of \micro, a code that calculates the relic density of the lightest supersymmetric particle (LSP) in the minimal
 supersymmetric standard model (MSSM). 
 All tree-level processes for the annihilation of the
 LSP are included as well as all possible
 coannihilation processes with  neutralinos, charginos, sleptons,  squarks and gluinos. 
 The cross-sections extracted from {\tt CalcHEP} are calculated exactly using 
  loop-corrected masses and mixings as specified in the {\it SUSY Les Houches Accord}.  Relativistic formulae for the thermal average are used and care is taken to handle poles and thresholds by adopting specific integration routines.
  The input parameters can be either the soft SUSY parameters in a general MSSM
 or the  parameters of a SUGRA model specified at some high scale (GUT). In the latter case, a link with \suspect, \softsusy, \spheno~ and \isajet~  allows to
 calculate the supersymmetric spectrum, Higgs masses, as well as mixing matrices.
 Higher-order corrections to Higgs couplings to quark pairs  
 including QCD as well as some SUSY corrections ($\dMb$) are implemented.  
  Routines calculating $(g-2)_\mu$, 
 $b\to s\gamma$ and $\bsmu$ are also included.
 In particular the $\bsgamma$ routine includes an improved NLO for the SM 
 and the charged Higgs while the SUSY large $\tan\beta$ effects beyond leading-order are included.
 This new version also provides 
 cross-sections for any $2\to 2$ process as well as partial decay widths for two-body final states in the MSSM allowing for easy simulation at colliders.}
\vspace*{\fill}


\vspace*{0.1cm}
\rightline{LAPTH-1044}
\rightline{{\today}}
\end{titlepage}

\setcounter{section}{0}
\setcounter{subsection}{0}
\setcounter{equation}{0}

\def\thesubsection {\thesection.\arabic{subsection}}
\def\theequation{\thesection.\arabic{equation}}

\baselineskip=18pt

\section{Introduction}

We present  \micronew, a program which  calculates the relic density of the lightest supersymmetric particle (LSP) in the minimal supersymmetric standard model (MSSM).
The stable LSP, which occurs in supersymmetric models with R parity conservation, constitutes a good candidate for cold dark matter.
Recent measurements from WMAP\cite{wmap} have in fact constrained the value
for the relic density  within 10\%, 
$$.094<  \Omega h^2<.128 \;\;\;\;{\rm at} \;\; 2\sigma.$$
Forthcoming experiments by the PLANCK satellite\cite{planck} will pin-down
this important parameter to within 2\%.
One therefore needs to evaluate the relic density with high accuracy.

The relic density calculation is based on solving the  equation characterizing the  evolution of the number density of the LSP.
For this, one needs to evaluate the thermally averaged cross-section for annihilation of the LSP,
 as well as, when necessary, coannihilation with other supersymmetric (SUSY) particles \cite{coannihilation,coannihilation2,Gomez}.
 We use, as in \micro1.1 \cite{belanger_cpc}, the method described in \cite{GondoloGelmini}
for the relativistic treatment of the thermally averaged
cross-section, and the generalization of \cite{EdsjoGondolo} to the case
of coannihilation. 
However we have improved our method for solving
the density evolution  equation, it is now solved
numerically without using the freeze-out approximation.
This improvement has not impaired the speed of the calculation.

The other main improvement to \micronew~ is the use of  loop corrected  
superparticle masses and mixing matrices.
These masses and mixing matrices, as specified in the {\it SUSY Les Houches Accord} (SLHA)\cite{Skands},
are then used to compute exactly all annihilation/coannihilation cross-sections.
This can be done whether the input parameters are specified at the weak scale
or at the GUT scale in the context of SUGRA models or the like. In the last
case, loop corrections are obtained from one of the public codes
which  calculate the supersymmetric spectrum using renormalization
group equations (RGE) \cite{SUSPECT,SOFTSUSY,SPHENO,ISAJET}.
 These corrections to masses are critical
for a precise computation of the relic density in two specific
regions: the coannihilation region and the region where
annihilation through a Higgs or Z exchange occurs near resonance.
Note that these regions of the supersymmetric parameter space are among the ones where one  predicts  sufficiently high annihilation rates for the neutralino
LSP to meet the WMAP upper bound on the relic density.
In the first case, the critical parameter is the NLSP-LSP mass
difference, in the latter the mass difference $2M_{LSP}-m_{H/Z}$ \cite{houches_compar}.
The Higgs masses are calculated either by one of the RGE codes
or  with \feynhiggs\cite{FeynHiggs}.
When annihilation occurs near a Higgs resonance, higher-order
corrections to the width also need to be taken into account. As in
\micro1.1, QCD corrections to Higgs partial widths are included,
furthermore we have added the important SUSY corrections, the $\dMb$ correction, that are
relevant at large $\tan\beta$. These higher-order corrections also  
affect directly the Higgs-$q\overline{q}$ vertices and are taken into account 
in all the relevant annihilation cross-sections.

Besides the relic density measurement, other direct or indirect precision measurements
constrain the supersymmetric models. In our package we calculate the
supersymmetric contribution to $\gmuon$ and to $\delrho$.
We also include a new  calculation of the supersymmetric contribution to
$\bsmu$ and an improved calculation
of the $\bsgamma$ decay rate. The latter includes an improved NLO for the SM and the charged Higgs contribution as well as the SUSY large $\tan\beta$ effects
beyond leading-order,  the  $\dMb$ correction.  The $\bsgamma$, $\bsmu$
or $\gmuon$ routines can be  replaced or used as a stand-alone code.

Within \micronew~ all (co-)annihilation cross-sections are compiled by 
\calchep~\cite{CalcHEP} which is included in the package. 
\calchep~ is an automatic program to calculate tree-level cross-sections
for any process in a given model, here the MSSM.
We provide a code that performs the calculation of cross-sections and decay widths that can be called  independently of the relic density calculation.
The input parameters are  the parameters of the {\it SUSY Les Houches Accord}.
Another  new feature is the possibility to call \calchep, directly from a \micronew~ session, and
calculate interactively cross-sections for any process 
in the MSSM or in mSUGRA models.  For this, all widths of supersymmetric particles are evaluated automatically at tree-level, including  the available two-body decay modes. The relic density  as well as other constraints are also calculated in the \calchep~ session.

In summary, the new program \micronew
\begin {itemize}
\item{} Calculates complete tree-level matrix elements for all subprocesses.
\item{} Includes all coannihilation channels, in particular channels with neutralinos, charginos,
sleptons, squarks and gluinos.
\item{} Calculates the relic density for any LSP, not necessarily the lightest neutralino.
\item{} $^*$Deals with two sets of input parameters:   parameters of the MSSM understood to be specified at the EWSB scale or parameters of the SUGRA model specified at the GUT scale.
Both mSUGRA or non-universal SUGRA models are included.
\item{} $^*$Provides an interface with the main codes to calculate the supersymmetric spectrum:
\suspect~\cite{SUSPECT}, \isajet~\cite{ISAJET}, \spheno~\cite{SPHENO} and \softsusy~\cite{SOFTSUSY}. 
\item{} $^*$Includes an interface  with the {\it SUSY Les Houches Accord} \cite{Skands} for supersymmetric model specifications and input parameters.
This gives a lot of flexibility as any model for which the MSSM
spectrum is calculated by an external code can be incorporated easily.
\item{} $^*$Includes loop corrected sparticle masses and mixing matrices.
\item{} $^*$Includes loop-corrected Higgs masses and widths.
 QCD corrections
to the Higgs couplings to  fermion pairs are included as well
as, via an effective Lagrangian,  the $\dMb$ correction relevant at large $\tan\beta$.
\item{} $^*$Provides exact numerical solution of the Boltzmann equation by Runge-Kutta.
\item{} Outputs the relative contribution of each channel to $1/\Omega$
\item{} $^*$Computes cross-sections for any $2\ra 2$ process
at the parton level.
\item{} $^*$Calculates decay widths for all  particles at tree-level
including all $1\ra 2$ decay modes.
 \item{} $^*$Calculates NLO corrections to $\bsgamma$.
\item{} $^*$Calculates constraints on MSSM: $\gmuon$, $\delrho$, $\bsmu$.
\item{} $^*$Supports both C and {\tt Fortran}.
\item{} Performs rapidly the relic density calculation, the limiting factor in the execution time of the program is the computation of the supersymmetric spectrum.
\end{itemize}

New features in the list above are denoted by a star.
In this paper we emphasize mainly the new features of our package,
full details can be found in the original reference~\cite{belanger_cpc}.
In Section 2, we describe the main changes to our calculation of the relic
density. We then give the parameters of the supersymmetric model
used in our package.
A description of the package follows in Section 4. Section 5 gives instructions for running the program as well as sample sessions. Finally in Section 6 we compare our results with those of \darksusy4.0 \cite{Darksusy}, the other public package that computes the relic density of supersymmetric dark matter.

\section{Calculation of the relic density}
\label{solution}

The most complete
formulae for the calculation of the abundance $Y(T)$ were presented in
\cite{GondoloGelmini,EdsjoGondolo} and we will follow their approach rather
closely.
The evolution equation for the abundance, defined as the number density divided by the entropy density,
  writes
\begin{equation}
 \frac{dY}{dT}= \sqrt{\frac{\pi  g_*(T) }{45}} M_p <\sigma v>(Y(T)^2-Y_{eq}(T)^2)
    \label{dydt}
\end{equation}
where $g_{*}$ is an effective number of degree of freedom \cite{GondoloGelmini}, $M_p$ is the Planck mass and
$Y_{eq}(T)$ the thermal equilibrium abundance.
$<\sigma v>$ is the relativistic thermally averaged
annihilation cross-section of  superparticles summed over all channels,
\begin{equation}
       <\sigma v>=  \frac{ \sum\limits_{i,j}g_i g_j  \int\limits_{(m_i+m_j)^2} ds\sqrt{s}
K_1(\sqrt{s}/T) p_{ij}^2\sigma_{ij}(s)}
                         {2T\big(\sum\limits_i g_i m_i^2 K_2(m_i/T)\big)^2 }\;,
\label{sigmav}
\end{equation}
where $g_i$ is the number of degree of freedom,  $\sigma_{ij}$  the total cross section for annihilation of
a pair of supersymmetric particles with masses $m_i$, $m_j$ into some Standard 
Model particles, and  $p_{ij}(\sqrt{s})$ is the momentum (total energy) of the incoming
particles in their center-of-mass frame.

 Integrating Eq.~\ref{dydt} from $T=\infty$ to
$T=T_0$  leads to the present day  abundance $Y(T_0)$
 needed in the estimation of the relic density,
\beqn
\label{omegah}
\Omega_{LSP} h^2= \frac{8 \pi}{3} \frac{s(T_0)}{M_p^2 (100{\rm(km/s/Mpc)})^2} M_{LSP}Y(T_0)= 
 2.742 \times 10^8 \frac{M_{LSP}}{GeV} Y(T_0)
\eeqn
where $s(T_0)$ is the entropy density at present time and $h$ the normalized Hubble constant. The present-day energy density is then simply expressed as
$\rho_{LSP}=10.54 \Omega h^2 (GeV/m^3)$.

Let us rewrite Eq.~\ref{dydt} in terms of $X=T/M_{LSP}$
\begin{equation}
   \frac{dY}{dX}= A(X) (Y_{eq}(X)^2- Y(X)^2) 
    \label{dydx}
\end{equation}
\begin{equation}
   A(X)= \frac{M_{LSP}}{X^2}\sqrt{\frac{\pi  g_*(M_{LSP}/X) }{45}} M_p <\sigma v>                                                                             
\label{Acoef}
\end{equation}

First note that  one will always 
have $Y(X) \approx Y_{eq}(X)$ when   $ A(X) Y_{eq}(X) \gg 1$.
This is  the case at $X \leq 1$ since the equilibrium abundance  $Y_{eq}(X)\approx{\cal O }(1)$ \cite{GondoloGelmini,EdsjoGondolo} and
for a typical electroweak cross-section ,  $<\sigma v>\approx10^{-10} GeV^{-2}$, and LSP mass,
$M_{LSP} \approx 100 GeV$ , one has  $A\approx 10^{10}$.
Choosing a starting point for the solution of the numerical equation
at small $X$ will rapidly  return the solution $Y=Y_{eq}$. 
On the other hand when $X>1$, $Y_{eq}(X)$  decreases exponentially as  
$e^{-X}$. 
Then  neglecting the dependence on X in both $A(X)$ and $Y_{eq}(X)e^{X}$   we
get 
\begin{equation}
   \Delta Y=Y(X)-Y_{eq}(X)=\frac{1}{2A}
\label{dy}
\end{equation}
where  $\Delta Y \ll Y_{eq}$. In this approximation,  $\Delta Y$
does not depend on $X$, whereas $Y_{eq}(X)$ decreases exponentially.  
This can be used 
to find a starting point $X_{f_1}$ for  the numerical solution of the differential  equation (\ref{dydx}). In order to find  $X_{f_1}$ where  
$\Delta Y(X_{f_1})=\delta \;\; Y_{eq}(X_{f_1})$ 
one can  solve
\begin{equation}   
 Y_{eq}(X_{f_1})'=A(X_{f_1})*Y_{eq}(X_{f_1})^2\delta (\delta+2)
 \label{fzot}
\end{equation}

In the \verb|darkOmega| routine we use this equation to find 
$X_{f_1}$, $Y(X_{f_1})$ corresponding to $\delta=0.1$ and solve the  differential
equation (\ref{dydx})  by the Runge-Kutta method starting from this point
\cite{NumRec}. We stop the  Runge-Kutta run at point $X_{f_2}$ where   
\begin{equation}
\label{Xf2}
Y_{eq}(X_{f_2})< \frac{1}{10} Y(X_{f_2})\;\;\; .
\end{equation}
Then we integrate Eq.~\ref{dydx} neglecting 
 the term $Y_{eq}(X)$
\beqn
\label{fzot2}
\frac{1}{Y(X_0)}=\frac{1}{Y(X_{f_2})}+\int_{X_{f_2}}^{X_0}
A(X) dX\;\;. 
\eeqn
Note that the  temperature  $T_0=2.725K$ corresponds to $X_0 \approx 10^{14}$.
Thus without loss of precision we can set  $X_0=\infty$  for evaluating $Y_0$
since $A(X) \propto 1/X^2$.

Another routine \verb|darkOmegaFO| performs  the calculation in
the {\it freeze-out} approximation{\footnote{This function was used in the original version of \micro \cite{belanger_cpc}.}. Here we choose $\delta=1.5$  as in  
 Ref.\cite{GondoloGelmini} and omit the  Runge-Kutta step 
($X_f=X_{f_1}=X_{f_2}$). The  precision of this
approximation is about 2\% although in  some exotic cases the  
approximation works badly.

As in \micro1.1, we include in the thermally averaged 
 cross-section , Eq.~\ref{sigmav}, only the contribution 
 of processes for which the  Boltzmann
suppression factor, $B$,  is above some value $B_\epsilon$  
\begin{equation}
\label{beps}
B=\frac{K_1((m_i+m_j)/T)}{K_1(2m_{LSP}/T)}\approx
e^{-X\frac{(m_{i}+m_{j}-2m_{LSP})}{m_{LSP}}} > B_\epsilon
\end{equation}
where $m_{i},m_{j}$ are the masses of the incoming superparticles.
The recommended value is $B_\epsilon=10^{-6}$\cite{belanger_cpc}.

 In our program we  provide two options to do the integrations,
the {\it fast} one and the {\it accurate} one. The  {\it fast} mode already gives
a   precision of about 1\% which is good enough for all practical purposes.
The  {\it accurate}  mode should be used only for some checks.
In the  {\it accurate}  mode   the program evaluates all integrals by means of
an adaptative Simpson  program. It automatically detects all singularities of
the integrand and  checks the precision.
In the case of the {\it fast} mode the accuracy  is not checked. We integrate the
squared matrix elements over  the scattering angle by means of 
a 5 points Gauss formula. For integration over $s$, Eq.~\ref{sigmav},
 we use a restricted set
of points which depends whether we are in the vicinity of
a s-channel  Higgs/Z/W resonance or not. We increase the number of points if
the Boltzmann factor corresponding to  $m_{pole}$
is larger  than $ 0.01 B_\epsilon$.

\subsection{Decays of the Higgs scalars}
\label{Higgs_decay}
When the LSP is near a Higgs resonance, it annihilates very efficiently.
The value of the neutralino annihilation cross-section  depends on the total
width if this width is larger than $\approx T_f/X_f$, the freeze-out temperature.
This is usually the case for large Higgs masses of 1TeV especialy 
at large $\tan\beta$ due to the enhancement in  the $\bbar$ channel. However
the width of $h(H,A)\ra b\bar{b}$ receives important QCD corrections. Typically
for the heavy Higgses ($m_H>1$TeV) the partial width into $q\bar{q}$ can vary easily by a
factor of 2 from the tree-level prediction, due mostly to the running of the quark mass at high scale.
To take these corrections into account we have redefined the vertices
 $hq\overline{q},Hq\overline{q}$ and $Aq\overline{q}$   using an effective mass
that reproduces the radiatively corrected Higgs  decays \cite{HDECAY}.
The effective mass at the scale $Q$ writes
\begin{equation}
\label{meff}
 M_{eff}^2(Q)=M(Q)^2\left[1+5.67a + (35.94-1.36n_f)a^2 +
    (164.14-n_f(25.77-0.259n_f))a^3\right]
\end{equation}
where
$a=\alpha_s(Q)/\pi$, the scale of the reaction is set to  $Q=2\mneuto$,   $M(Q)$  and $\alpha_s(Q)$    are the quark masses and
running strong coupling  in the $\overline{MS}$-scheme.
We use NNLO expressions for the strong coupling constants \cite{PDG}
and  for the running quark masses \cite{HDECAY,melnikov}.
The relation between the $\overline{MS}$ and the pole quark masses
are implemented  at three-loops \cite{PDG, melnikov}.
These are relevant for the top quark, since we use the pole mass as input
following the {\it SUSY Les Houches Accord} \cite{Skands}.
For b-quark, although $\mb(\mb)^{\overline{MS}}$ is the input parameter, it is still necessary to compute the pole mass used as an input parameter to some of the RGE codes.
 We set  $M_{eff}(Q)=M_{pole}$ at scales where the
effective mass  exceeds the  value of the pole mass.

We also take into account  the SUSY-QCD corrections
\cite{Guasch:2003cv} to $h,H,A \to b\bar{b}$ vertices that are
important at  large $\tan\beta$. Here  we use the effective
Lagrangian

\begin{eqnarray}
{\cal L}_{eff}= \sqrt{4\pi\alpha}_{QED} \frac{\mb}{1+\dMb}\frac{1}{2 \mw\sw}
\left[ - H b\bar{b}\frac{\cos\alpha}{\cos\beta}
\left(1+\frac{\dMb \tan\alpha}{\tan\beta}\right)\right.\nonumber\\
+ i Ab\bar{b}
 \tan\beta\left(1-\frac{\dMb}{\tan\beta^2}\right)
+h b\bar{b}\left.
\frac{1}{\cos\beta}
\left (1-\frac{\dMb}{\tan\alpha\tan\beta}\right)\right]
 \end{eqnarray}

\noi
where $\mb$ is the effective b-quark mass described above,
$\alpha_{QED}$ the electromagnetic coupling,
$\tan\beta$ is the ratio of the vev's of the Higgs doublets  and $\alpha$ is the Higgs mixing angle.
$\dMb$ is a correction factor arising from loop contribution of SUSY particles.
This factor is particularly important at large $\tan\beta$ and also
contributes to  $\bsgamma$ (all details are given in  Appendix B).

In the large $\tan\beta$ case, when
neutralino annihilation via s-channel Higgs exchange  dominates,
the inclusion of SUSY-QCD corrections can shift by about  15\% the value for the relic density. There is an option to switch off this correction (see Section\ref{assignment}).

The total width of the Higgs includes only the two-body final
states that occur at tree-level. In the case of the light Higgs,
this underestimates the width since the partial width to off-shell
W or  $gg$ final states can reach 10\%. However an accurate value
for this very narrow width has in general not a strong  impact on
the relic density. On the other hand a precise value for the heavy
Higgs width is necessary.
\def\o1{\chi^0_1}
\def\se1{\tilde{e}}

\subsection{Neutralino ``width"}
\label{Neutralino_width}

We assume that the LSP  is stable because of R-parity
conservation, however it is necessary to introduce a width  for
this stable particle in order to avoid infinities in some
processes. For example, in the coannihilation process like
$\sel\o1\ra eX$ via t-channel exchange of $\o1$ an infinity is
caused by the pole in the propagator, this is due to the fact that one can have 
a  real  decay  $\sel\ra e\neuto$.
  We assign 
a value of $\verb|sWidth|\cdot M_{LSP}$ to the  width of all supersymmetric particles .
The default value for the variable $\verb|sWidth|$ is 0.01.

\subsection{ Loop corrections to the MSSM spectrum.}

In the mSUGRA model, but also in the more general MSSM,
annihilation of the LSPs  near a Higgs or Z resonance and/or coannihilation
processes are often the dominant reactions in models where
$\Omega h^2 \approx 0.1$ \cite{sugra}.
For an accurate calculation of  the relic density it is then very important to have the exact relations between particle masses.
In particular,  the direct annihilation of a pair of neutralinos ($\neuto$)  depends sensitively
on the mass difference with the  Higgs or Z, $2\mneuto-M_{H/Z}$,  when the annihilation occurs near the resonance. Furthermore
coannihilation processes depend strongly on  the NLSP-LSP  mass difference.

In this new version of \micronew, we provide an option to
calculate loop corrections to all sparticle masses {\footnote{Pole masses in the calculation of the relic density were first used in Ref.~\cite{baerbelyaev}}}. Within the
MSSM defined at the EWSB scale, loop corrections are implemented by a call to \suspect\cite{SUSPECT},
within the SUGRA or other model defined at the GUT scale, the loop
corrections are done by any of the four public  codes (\suspect~,\softsusy~,
\spheno~,\isajet~) for
calculating the supersymmetric spectrum based on renormalization
group equations. Because  it is a mass difference rather than the
absolute mass that has a large impact on the prediction of the
relic density, even radiative corrections at the percent level,
such as is often the case for neutralinos, need to be taken into
account. Indeed large shifts in the prediction of the relic
density between tree-level and loop-corrected masses can be found.
Typically the prediction for the relic density can change by
20\%, but in some scenarios corrections  can reach 100\% or even
more. We not only use the loop-corrected sparticle masses but also
the corresponding mixing matrix elements. In this way we take into
account some of the loop corrections in the evaluation of the
matrix elements for different processes. This however means, since
it is only a partial implementation of loop corrections, that
theoretical inconsistencies in the model could occur, in
particular
 problems with unitarity violation in some processes. This would mainly show up
 in processes with production of gauge particles,  however at much higher energies
 that are typically involved in the LSP annihilation processes.

\section{The MSSM parameters.}

In our package, we compute  various matrix elements and cross-sections for
$2\ra 2$ processes within the framework of the MSSM.
The model file corresponding to the specific  implementation of the MSSM was
derived with LanHEP\cite{Lanhep}, a program that
generates the complete set of particles and vertices once given a  Lagrangian
\cite{Semenov:2002mssm,semenov_higgs}.
Names are attached to the parameters
of the MSSM, including those of the SM,  and their values can be set
with an instruction. For example, the command  \verb| assignVal("Mtp",180.)|  assigns the
value $m_t=180$GeV to the pole mass of the t-quark.



The list of  parameters of the Standard Model and their default values is
presented in Table~\ref{SMpar}.
All quarks and leptons of the  first two generations are assumed
 massless.
 The default values for the  electromagnetic coupling and  the Weinberg angle
correspond to the values in the  $\overline{MS}$  scheme at the $\mz$ scale.

\begin{table*}
\caption{\label{SMpar} Standard Model parameters}
\vspace{.5cm}
\begin{tabular}{|l|l|l|}
\hline
name      &default      & definition\\
\hline
AlfEMZ    &0.00781653    & electromagnetic coupling $\alpha_{em}(M_Z)$ \\
AlfSMZ    &0.1172       &  strong coupling, $\alpha_s^{\overline{MS}}(M_Z)$ for $n_f=5$   \\
SW        &0.481        & Weinberg angle, $\sin\theta_W$  \\
MZ        &91.1884      & Z mass\\
Ml        &1.777        &tau-lepton pole mass \\
Mtp       &175.0        & t-quark pole mass \\
MbMb      &4.23         &$\overline{MS}$ scale independent b-quark mass Mb(Mb)\\
\hline
\end{tabular}
\end{table*}

The parameters of the  MSSM  are described in  Table~\ref{LesHouchesPar}.
We follow the conventions of the {\it SUSY Les Houches Accord}
\cite{Skands}. The masses of the third generation fermions are  ordered,
for example $\msto$ corresponds to the lightest top-squark.
In this list, the number of parameters exceeds the number
of MSSM independent parameters. They correspond to physical parameters, masses and mixings. This extended set of parameters is however necessary when one wants to use effective masses and vertices that include loop corrections.
Our computation of matrix elements for cross-sections is based on this set of parameters.
Note that the trilinear muon coupling, $A_\mu$, is added to the parameter list even though it does not contribute to matrix elements or to the  spectrum since the muon is assumed to be massless. This parameter is however important  for evaluating the muon anomalous  magnetic moment.

\begin{table*}[htb]
\caption{\label{LesHouchesPar}
MSSM parameters of the SUSY Les Houches Accord}
\vspace{.5cm}
\begin{tabular}{|l|l||l|l|}
\hline
name     & comment                &                  name          & comment                          \\
\hline
tb    & $\tan\beta$      &             MSnl     &$\tau$-sneutrino mass\\
alpha    & Higgs $\alpha$ angle  &              MSe$\frac{L}{R}$ & masses of  left/right selectrons \\
mu   & Higgs $\mu$ parameter  &             MSm$\frac{L}{R}$ &left/right smuon masses        \\
Mh   & Mass of light Higgs   &              MSli   &i=1,2 masses of   light/heavy $\tilde{\tau}$ \\
MH3  & Mass of CP-odd Higgs  &              MSu$\frac{L}{R}$   &  masses of left/right u-squarks  \\
MHH  & Mass of Heavy Higgs   &              MSs$\frac{L}{R}$    & masses of left/right s-squarks \\
MHc  & Mass of charged Higgs &              MSti& i=1,2 masses of  light/heavy  t-squarks   \\
Al   & $\tilde{\tau}$ trilinear coupling  & MSd$\frac{L}{R}$& masses of  left/right d-squarks\\
Am       & $\tilde{\mu}$ trilinear coupling   & MSc$\frac{L}{R}$& masses of  left/right c-squarks   \\
Ab   & $\tilde{b}$ trilinear coupling     & MSbi   &i=1,2 masses of   light/heavy b-squarks \\
At   & $\tilde{t}$ trilinear coupling     & Zn$_{ij}$   & i,j=1,..,4;  neutralino mixing matrix\\
MNEi  & i=1,2,3,4; neutralino masses          & Zu$_{ij}$   & i=1,2;j=1,2; chargino U mixing matrix \\
MCi & i=1,2 chargino masses & Zv$_{ij}$   & i=1,2;j=1,2; chargino V mixing matrix \\
MSG  & mass of gluino             & Zl$_{ij}$   & i=1,2;j=1,2; $\tilde{\tau}$ mixing matrix \\
MSne     & $e$-sneutrino mass             & Zt$_{ij}$   & i=1,2;j=1,2; $\tilde{t}$ mixing matrix\\
MSnm     &$\mu$-sneutrino mass            & Zb$_{ij}$   & i=1,2;j=1,2; $\tilde{b}$  mixing matrix\\
\hline
\end{tabular}
\end{table*}

The values of the SLHA parameters can
either be set by an external program, here a call to one of the
RGE codes that calculate the
 supersymmetric spectrum, or by specifying the  MSSM parameters at the weak scale.
 In either case one needs to  specify a set of {\it independent} parameters
 as described below.

\subsection{Input parameters at the GUT scale}

Within the context of the SUGRA scenario for supersymmetry
breaking the MSSM parameters can be evaluated at the weak scale
starting from a set of scalar masses, gaugino masses, trilinear
couplings defined at the GUT scale.  The GUT scale input
parameters are listed in Table~\ref{standardMSSM}. Only one parameter,
$\tan\beta$, is defined at $\mz$. We implicitly assume that the
first two generations are identical. The parameters for the mass of the Higgs doublet can be entered with a negative sign, in this case they  will be understood as
$M_{H_U}^2=- |M_{H_U}|^2$

We treat the mSUGRA model as a special case of the general SUGRA.
Since  simplifying relations are  imposed on   masses and
couplings, in the mSUGRA model one has to specify only a small
number of input parameters  at the GUT scale: $\m0,\mhf,A_0,
\tan\beta, sgn(mu)$.
These correspond to  \\
\hspace*{1cm} {\tt m0} = $Mli=Mri=Mqi=Mui=Mdi=MHu=MHd$ \\
\hspace*{2cm}- common scalar mass at GUT scale;\\
\hspace*{1cm} {\tt mhf}= $MG1=MG2=MG3$ -   common gaugino mass at GUT scale;\\
\hspace*{1cm} {\tt a0}=$At=Ab=Al$ - trilinear soft breaking parameter at GUT scale;\\
\hspace*{1cm} {\tt tb}-      $\tan{\beta}$ or the  ratio of vacuum expectation values at MZ;\\
\hspace*{1cm} {\tt sgn}-     +/-1,  sign of $\mu$, the Higgsino mass term.\\

Four different routines  read  the parameters
of  Table~\ref{LesHouchesPar} and pass them to the corresponding  packages  that solves the RGE equations and calculate the MSSM masses and mixing matrices. The routines
\verb|suspectSUGRA| \cite{SUSPECT},\verb|softsusySUGRA| \cite{SOFTSUSY},
\verb|sphenoSUGRA| \cite{SPHENO},
\verb|isajetSUGRA| \cite{ISAJET}
 are described in section\ref{assignment}.
Note that some of the standard parameters of Table~\ref{SMpar} also
play a role in the  low energy boundary conditions implemented in the RGE codes.  They are passed to RGE routines implicitly.
We assume that
the second generation is identical to the first one and only parameters of the first generation are used.

\begin{table*}[htb]
\caption{\label{standardMSSM} Independent GUT-scale parameters.}
\vspace{.5cm}
\begin{tabular}{|l|l||l|l|}
\hline
name& comment &name& comment\\
\hline
tb  &  $\tan\beta$ (at $\mz$) &  Ml1 & Left-handed slepton mass for $1^{st}/2^{nd}$ gen. \\
At  & $\tilde{t}$ trilinear coupling&     Ml3 & Left-handed slepton mass for  $3^{rd}$ gen. \\
Ab  & $\tilde{b}$ trilinear coupling&      Mr1 & Right-handed slepton mass for  $1^{st}/2^{nd}$ gen.\\
Al  & $\tilde{\tau}$ trilinear coupling& Mr3 & Left-handed slepton mass for  $3^{rd}$ gen. \\  
MG1 & U(1) Gaugino mass&    Mq1 & Left-handed squark mass for  $1^{st}/2^{nd}$ gen. \\
MG2 & SU(2) Gaugino mass& Mq3 & Left-handed squark mass for  $3^{rd}$ gen. \\ 
MG3 & SU(3) Gaugino mass & Mu1 & Right-handed u-squark mass  for $1^{st}/2^{nd}$ gen. \\
sgn & sign of $\mu$ at the EWSB scale &Mu3 & Right-handed u-squark mass  for $3^{rd}$ gen. \\
MHu & Mass of first Higgs doublet& Md1 & Right-handed d-squark mass for $1^{st}/2^{nd}$ gen. \\
MHd & Mass of second Higgs doublet& Md3 & Right-handed d-squark mass for $3^{rd}$ gen. \\
\hline
\end{tabular}
\end{table*}

\subsection{Input parameters at the weak scale}

The parameters of the  {\it SUSY Les Houches Accord} can also be
calculated starting from the set of {\it independent} MSSM
parameters at the EWSB scale\footnote{ This set of parameters was used in the
previous version of \micro\cite{belanger_cpc}.}
 listed in Table~\ref{standardMSSM}\cite{Semenov:2002mssm}. This can
be done either at tree-level or with loop corrections (see
Section~\ref{assignment}). The names of the independent parameters
of the MSSM are identical to the GUT scale parameters safe for
{\tt MHu,MHd} which are conveniently replaced by $\mu$ and $\ma$(MH3).
Furthermore at the EWSB scale one must define the sfermion masses for all three generations.
 Here MH3 and MG3 are the pole
masses of the CP-odd Higgs and of the gluino. All other parameters
are treated as running ones.
 When  evaluating loop corrections to  pole masses starting from the independent set of parameters,
it is assumed that the parameters are specified in the
$\overline{DR}$ scheme  at the EWSB scale,  $Q=\sqrt{\msto\cdot
\mstt}$.
\begin{table*}[htb]
\caption{\label{ewsbMSSM} Set of independent MSSM parameters
at the weak scale.} \vspace{.5cm}
\begin{tabular}{|l|l||l|l|}
\hline
name& comment &name& comment\\
\hline
tb  &  $\tan\beta$ &                      MG3 & SU(3) Gaugino mass (gluino mass)\\
mu  & Higgs $\mu$ parameter &              Mli & Left-handed slepton mass for  $i^{th}$ generation \\
At  & $\tilde{t}$ trilinear coupling&      Mri & Right-handed selectron mass for $i^{th}$ generation \\
Ab  & $\tilde{b}$ trilinear coupling&      Mqi & Left-handed squark mass for $i^{th}$ generation \\
Al  & $\tilde{\tau}$ trilinear coupling&   Mui & Right-handed u-squark mass  for $i^{th}$ generation \\
Am  & $\tilde{\mu}$ trilinear coupling&    Mdi & Right-handed d-squark mass for $i^{th}$ generation \\
MG1 & U(1) Gaugino mass&                   MH3 & Mass of Pseudoscalar Higgs \\
MG2 & SU(2) Gaugino mass& & \\
\hline
\end{tabular}
\end{table*}

Two options are available to specify  the weak scale MSSM parameters, either from a file using the function \verb|ewsbInitFile| or directly as argument of the function \verb|ewsbMSSM|.
Either option will evaluate the supersymmetric spectrum at tree-level or to one-loop 
according to the value of the parameter \verb|LCOn|, see section~\ref{assignment}.

After evaluation of the spectrum in the context of the SUGRA or MSSM models,  the function \verb|calcDep| chooses the lightest supersymmetric particles
and calculates the running masses of quarks at the LSP scale as well as various widths.

\section{Functions of \micro}

The routines presented below belong to  the {\tt micromegas.a}
library. They are available both in  the C and Fortran versions.
If for some reason a  Fortran call differs from the C one, we
present the Fortran version in
 brackets "[~]".   
 The types of  the functions and their arguments are specified in
Appendix~A.
 Examples of  implementation are presented in Section \ref{examples}.
Note that after assignments of the  MSSM parameters the user has to call the initialization procedure \verb|calcDep| (Sec.~\ref{assignment}). Other routines of the
package can only be used after making this call.

\subsection{Variable assignment and spectrum calculation}
\label{assignment}
\noi$\bullet$\verb|assignVal(name,val)|\\
changes  values of the parameters.
{\tt name} is  one of  the names  presented in
Tables~\ref{SMpar},\ref{LesHouchesPar},
{\tt val} is the value to be assigned.
The function returns 0 when it successfully recognizes
the parameter name  and 1 otherwise. \\
\noi$\bullet$\verb|assignValW(name,val)|\\
the same routine as \verb|assignVal|, instead of returning an  error code it
writes a warning  on the screen.

\noi$\bullet$\verb|suspectSUGRA(tb,MG1,MG2,MG3,Al,At,Ab,sgn,MHu,MHd,Ml1,Ml3,Mr1,Mr3,Mq1,Mq3,|\\
\hspace*{3.1cm}\verb|Mu1,Mu3,Md1,Md3)|\\
calculates  the values of the MSSM parameters in the SUGRA
scenario using the \suspect\\  package. Returns $0$ when the
spectrum is  computed  succesfully, $1$ in case of non-fatal
problems (see the manual for  the meaning of non-fatal errors
\cite{SUSPECT}),  and $(-1)$  if no solution to RGE can be found
for a given set of boundary conditions. This routine assigns values
for the parameters in Table~\ref{LesHouchesPar}. The result  depends on the input values
of the SM parameters, in particular on the quark masses,
$\mt^{pole}, \mb(\mb)$ (\verb|Mtp|, \verb|MbMb|) and on the strong
coupling constant $\alpha_s(M_Z)$ (\verb|AlfSMZ|). These
parameters play a role in the low energy boundary conditions and
are passed implicitly.

\noi$\bullet$\verb|softSusySUGRA(tb,MG1,MG2,MG3,Al,At,Ab,sgn,MHu,MHd,Ml1,Ml3,Mr1,Mr3,Mq1,Mq3,|\\
\hspace*{3.1cm}\verb|Mu1,Mu3,Md1,Md3)|\\
same as above for \softsusy.

\noi$\bullet$\verb|sphenoSUGRA(tb,MG1,MG2,MG3,Al,At,Ab,sgn,MHu,MHd,Ml1,Ml3,Mr1,Mr3,Mq1,Mq3,|\\
\hspace*{3.1cm}\verb|Mu1,Mu3,Md1,Md3)|\\
 same as above for \spheno.

 \noi$\bullet$\verb|isajetSUGRA(tb,MG1,MG2,MG3,Al,At,Ab,sgn,MHu,MHd,Ml1,Ml3,Mr1,Mr3,Mq1,Mq3,|\\
\hspace*{3.1cm}\verb|Mu1,Mu3,Md1,Md3)|\\
same as above for \isajet. This function depends only on $m_t^{pole}$,
other SM parameters, and in particular  $\mb(\mb)$ and $\alpha_s$, are fixed internally. \isajet~ does not
calculate the trilinear muon coupling, we use the approximate relation for mSUGRA models, $A_\mu=A_0-0.7 M_{1/2}$.

Note  that only the \suspect~ code is included in our package.
Other codes should be installed independently by the user  and linked  to
\micro~ as  explained in  Section~\ref{INSTALL}.

\noi$\bullet$
\verb|ewsbMSSM(tb,MG1,MG2,MG3,Am,Al,At,Ab,MH3,mu,Ml1,Ml2,Ml3,Mr1,Mr2,Mr3,|\\
\hspace*{3.1cm}\verb|Mq1,Mq2,Mq3,Mu1,Mu2,Mu3,Md1,Md2,Md3,LCOn)|\\
calculates the supersymmetric spectrum at tree-level or one-loop 
from the set of independent MSSM parameters at the EWSB scale as specified by the parameter \verb|LCOn|. The Higgs sector parameters, masses and mixing
angle $\alpha$, are calculated with \feynhiggs~\cite{FeynHiggs}.

\verb|LCOn=0| - tree level formulae for super particles
masses;

\verb|LCOn=1| -  \suspect~ is used to
evaluate  loop corrections to masses of super particles. 

\noi$\bullet$\verb|ewsbInitFile(filename,LCOn)|\\
reads the input file {\tt filename} which specifies the set of independent MSSM parameters at the EWSB scale and calculates the supersymmetric spectrum
at tree-level or one-loop as set by the parameter  \verb|LCOn| (same as above).

The function returns: \\
\hspace*{1cm}  {\tt ~0} - when the input has been read correctly;\\
\hspace*{1cm}  {\tt -1} - if the file does not exist
or can not be opened for reading;\\
\hspace*{1cm}  {\tt -2} - if some parameter from Table~\ref{ewsbMSSM} is missing as displayed on the sceen;\\
\hspace*{1cm}  {\tt -3} - if the spectrum cannot be calculated;\\
\hspace*{1cm}  {\tt ~n} - when  the line number {\tt n} has been written in the
wrong format.\\
For example, the correct format of a line is \\
\hspace*{2cm} {\tt MG3 1500.} \\

\noi$\bullet$\verb|readLesH(filename,LE)|\\
reads the input file in the  {\it SUSY Les Houches Accord} format \cite{Skands}.
If \verb|LE=1| the SM parameters of Table 1 as well as
$\tan\beta$  are also read from a SLHA output file.

\noi$\bullet$\verb|calcDep(dMbOn)|\\
initializes  internal parameters for subsequent calculations. In
particular, the running masses of quarks, the strong coupling constant as well as the widths of gauge bosons, Higgses and superparticles.
Running parameters are evaluated at the LSP scale.
This routine also sorts the superparticles and selects the LSP.
The parameter \verb|dMbOn|$=0$ switches off SUSY-QCD corrections,
$\dMb$, see Section~\ref{Higgs_decay}.

\subsection{Display of parameters.}

\noi$\bullet$\verb|findVal(name,&val) [findVal(name,val) ]|\\
assigns to  the variable \verb|val| the value of the parameter
\verb|name|. It returns {\tt zero}
if such variable indeed exists and {\tt 1}  otherwise.
This function can be applied to any of the parameters in Table~\ref{SMpar},\ref{LesHouchesPar}
as well as to particle masses and widths specified in Tables~\ref{HiggsP},\ref{SSP},\ref{SMP}.\\
\noi$\bullet$\verb|findValW(name)|\\
returns the  value corresponding to the  variable \verb|name|. If
\verb|name| is not defined  \verb|findValW| writes a warning on the screen.\\
\noi$\bullet$\verb|printVar(file,N) [ printVar(N) ]|\\
prints  the first N  records of the full list of model parameters.
The first 7 parameters correspond to Table~\ref{SMpar}, the
following  75 parameters correspond to the list in
Table~\ref{LesHouchesPar}.  To see the parameters on
the screen,  substitute  {\tt file}={\tt stdout}.
In the Fortran version, only display on the sceen is possible. \\
\noi$\bullet$\verb|printMasses(file,sort) [ printMasses(sort) ]|\\
prints into the file the masses of the supersymmetric particles
as well as all Higgs masses and widths.
The  Fortran version writes down on the screen.
 If sort$\neq 0$, the masses are sorted in increasing  order.\\
\noi$\bullet$\verb|lsp() [ lsp(name) ]|\\
returns the name of the LSP.
The relic density can be calculated with any particle being the LSP even though
only the neutralino and the sneutrino can be dark matter candidates.
If the user wants to impose a specific LSP,  the nature of the LSP
must be checked after calling  {\tt calcDep}.

\noi$\bullet$\verb|lspmass_() [ lspmass() ] | \\
returns the mass of the lightest supersymmetric particle in $GeV$.

\subsection{ Calculation of relic density.}
\noi$\bullet$\verb|darkOmega(&Xf,fast,Beps) [ darkOmega(Xf,Fast,Beps) ]|\\
This is the basic function of the package
which returns the relic density $\Omega h^2$
(Eq.~\ref{omegah}).  The  procedure for solving  the evolution equation 
using Runge-Kutta was described in 
Section~\ref{solution}.
The  value of the freeze-out parameter {\tt Xf}  is returned
by the function and  equals  $(X_{f_1}+X_{f_2})/2$, (see the definition in  Eq.~\ref{fzot}, \ref{Xf2}).
The parameter {\tt Beps} defines the criteria for including a given channel
into the sum for the calculation of the thermally averaged cross-section,
Eq.~\ref{beps};  $10^{-6}$ is the  recommended value.

If {\tt fast=0},  we use an integration routine that increases the
number of points until an accuracy of $10^{-3}$  is reached. If
{\tt fast=1}  the accuracy is not checked, but a set of points is
chosen  according to the behaviour of the integrand: poles, thresholds, 
Boltzman suppression at large energy. The accuracy
of this mode is about 1\%. Finally, {\tt fast=2} corresponds to
the calculation of relic density  using the widely-used
approximation \cite{Gomez} based on the expansion in terms of velocity
$$ p\cdot\sigma(p)=A+B\cdot p^2.$$
 The recommended mode is {\tt fast=1}.

\noi
If some problem is encountered, {\tt darkOmega} returns $(-1)$.

\noi$\bullet$\verb|darkOmegaFO(&Xf,fast,Beps) [ darkOmegaFO(Xf,fast,Beps) ]|\\
calculates the relic density as the function \verb|darkOmega| described above, but using  the freeze-out approximation.

\noi$\bullet$\verb|printChannels(Xf,cut,Beps,prcnt,f) [ printChannels(Xf,cut,Beps,prcnt) ]|\\
prints the relative contribution to $\Omega^{-1}$  for all
subprocesses for which this contribution  exceeds the value chosen
for  {\tt cut}. If {\tt prcnt=1} the contribution is given in
percent, otherwise
 the absolute value is displayed.
It is assumed that the {\tt Xf} parameter was first
evaluated by {\tt darkOmega}. 
In the C version, the output is directed  to the file {\tt f},
the Fortran version writes on the screen.
Actually this routine  evaluates the partial contributions
to the integral of Eq.~\ref{fzot2} without the $1/Y_f$ term  and returns  
the corresponding value for $\Omega h^2$ .

\subsection{ Routines for constraints.}
\label{constrains}
\noi$\bullet$\verb|deltarho_() [ delrho()] |\\
calculates, by a call to a  \suspect~ routine,  the $\delrho$ parameter which
describes the MSSM corrections to electroweak observables. It contains
stop/sbottom contributions, as well as the two-loop QCD
corrections due to gluon exchange and the correction due to gluino
exchange in the heavy gluino limit \cite{deltarho}. Precise
measurements of SM electroweak observables allow to set the limit
$\delrho<2\cdot 10^{-3}$.

\noi$\bullet$\verb|bsgnlo_() [ bsgnlo() ]|\\
returns the value of the branching ratio for  $b\ra s\gamma$.
For $b\ra s\gamma$ we have improved on the results of \cite{bsgamma}
by including some very
recent new contributions beyond the leading order that are
especially important for high $\tan\beta$. Full details can be found in Appendix B.

\noi$\bullet$\verb|bsmumu_() [ bsmumu() ]|\\
returns the MSSM contribution to $\bsmu$.
Our calculation is based on
\cite{bsmumubobeth} and agrees with \cite{bsmumudreiner}.
It includes the loop contributions
due to chargino, sneutrino, stop and Higgs exchange. The $\dMb$ 
effect relevant for high $\tan \beta$ is taken into account.
The current bound from  CDF experiment at Fermilab is B.R.($\bsmu<9\times 10^{-6}$) \cite{CDFbsmumu} and the expected bound from RunIIa
should reach  B.R.($\bsmu<2\times 10^{-7}$) \cite{bsmu_leptonphoton}.

\noi$\bullet$\verb|gmuon_() [ gmuon() ]|\\
returns the value of the supersymmetric contribution to the
anomalous magnetic moment of the muon~\cite{g-2nous}. The
result depends only on  the parameters of the chargino/neutralino
sector as well as on the smuon parameters, in particular  the
trilinear coupling $A_\mu$ (\verb|Am|). Our formulas agree with
\cite{g-2_martin}. The latest experimental data on the $\gmuon$
measurement using $\mu^-$\cite{g-2ex}, brings the average to
$a_\mu^{\rm exp.}=11659208 \pm 6 \times 10^{-10}$. The quantity
$a_\mu$ includes both electroweak and hadronic contributions and
is still subject to large theoretical errors, the allowed range
for $\delta a_\mu=a_\mu^{\rm exp.} - a_\mu^{\rm theo.}$ then has
also large errors. We estimate the  $3\sigma$ range  to be $ 5.1
\;<\; \delta a_\mu\;\times 10^{10} \;<\; 64.1$ \cite{micro_sugra}.

\noi$\bullet$\verb|masslimits_() [ masslimits() ]|\\
returns a positive value  and
 prints a WARNING when the choice of parameters conflicts with a
direct accelerator limits on sparticle masses.
The constraint on the light Higgs mass is not implemented and must be
added by the user. 

Among the routines that calculate constraints, 
only \verb|masslimits| issues a warning  if the chosen model gives a value outside the experimentally allowed range.
All other constraints must be checked by the user.

\subsection{QCD auxiliary routines.}

\noi$\bullet$\verb| alphaQCD(Q)|\\ 
calculates the  running $\alpha_s$ at the scale \verb|Q| in the $\overline{MS}$ scheme.
The calculation is done using the \verb|NNLO| formula in 
\cite{PDG}. Thresholds for b-quark and t-quark  are
included in  $n_f$ at the scales $\mb(\mb)$ and $\mt(\mt)$ respectively. 
 Implicit input parameters are  \verb|AlfSMZ|, \verb|Mtp|, and
\verb|MbMb| defined in  Table 1.

\noi$\bullet$\verb| MbRun(Q), MtRun(Q) | \\
calculates top and bottom running masses evaluated at NNLO.

\noi$\bullet$\verb| MbEff(Q), MtEff(Q) | \\
calculates effective t- and b-quark masses as in  Eq.~\ref{meff}.

\noi$\bullet$\verb|deltaMb()|\\
calculates the SUSY corrections to $\dMb$ (Appendix B).

\subsection{Partial widths and cross sections}

\noi$\bullet$\verb| decay2(pName,k, out1, out2)|\\
calculates  the decay widths (in GeV) for any $1\rightarrow 2$  processes.
The input parameters are    {\it pName}, the name of the
decaying  particle and  {\it k}, 
the   channel number.  {\it out1} and 
{\it out2} are the names of outgoing particles  for channel  {\it k}.
If {\it k}  exceeds the total number of channels, then  
 {\it out1} and {\it out2} are filled as empty strings.

\noi$\bullet$\verb| newProcess(procName, libName) [ newProcess(procName, libName, address) ]|\\ 
 prepares  and compiles the  codes for any $2\ra 2$ reaction 
 in the MSSM.
The result of the compilation is stored in the library\\
\hspace*{3cm} \verb|source/2-2/|{\it libName}\verb|.os|.\\
 If this library already exists, it is not recompiled and the correspondence 
between the contents of the library and the {\it procName} parameter is not 
checked. 
{\it libName} is also attached to the names of routines in the {\it libName}.so library. Therefore {\it libName} should not  contain
 symbols such as $+,-,*,/,~$ which are not legal as identifiers.   
 Library names should not start with {\it omglib}, these are reserved 
 for the libraries  used to evaluate $\Omega h^2$.

The process should be specified in CalcHEP notations, for example \\
\hspace*{3cm} \verb| "e,E->~1+,~1-"|\\
without any blank space. One can find all symbols for MSSM  particles in  Tables~\ref{HiggsP},\ref{SSP},\ref{SMP}.
 Multi-process generation is also possible
using the command \\
\hspace*{3cm} \verb| "e,E->2*x"|\\
where \verb|x| means arbitrary  final states.

The \verb|newProcess| routine returns the  {\it address} of the static structure
with contains, for further use,  the code for the processes.  If the process can not
be compiled, then a NULL address is returned (address[1]=0 in Fortran).
\verb|newProcess| can  also return the address of a library that was already generated,
for example, \verb|newProcess("","omglib_o1_o1")| returns the address of the library for neutralino annihilation. 

\noi$\bullet$\verb|infor22(address,nsub,n1,n2,n3,n4,&m1,&m2,&m3,&m4)|\\
\verb|[ infor22(address,nsub,n1,n2,n3,n4,m1,m2,m3,m4) ] |\\
allows to check the contents of the library produced by 
\verb|newProcess|. Here {\it address} is the returned value of \verb|newProcess|
call and {\it nsub} the subprocess number. The  parameters returned correspond to the
names of particles for a given subprocess ({\it n1, n2, n3, n4}) 
as well as their masses ({\it m1, m2, m3, m4}).
The function returns 2 if the {\it nsub} parameters exceed the limits
and 0 otherwise.

\noi$\bullet$\verb| cs22(address, nsub, P, c1, c2 , &err)|\\
evaluates the cross section for a given $2\rightarrow 2$ process
with  center of mass momentum $P$(GeV). The differential cross section is integrated from  $ c1 < cos(\theta) <c2 $  and $\theta$ is 
the angle between $\bar{p}_1,\bar{p}_3$  in the center-of-mass frame.
If {\it nsub} exceeds the maximum value for the number of subprocesses  then  {\it err} contains a non zero error code.

\begin{table}
\caption{\label{HiggsP}Higgs particles.}
\vspace{.5cm}
\begin{tabular}{|l|l|l|l||l|l|l|l|}
\hline
 Name     &symbol&mass&width   &  Name     &symbol&mass&width    \\
\hline
Light Higgs  &h  &Mh    &wh    & CP-odd Higgs &H3 &MH3   &wH3    \\
Heavy higgs  &H  &MHH   &wHh   & Charged Higgs&H+,H- &MHc   &wHc \\
\hline
\end{tabular}
\end{table}

\begin{table}
\caption{\label{SSP}Names, masses and widths of
supersymmetric particles.}
\vspace{.5cm}
\begin{tabular}{|l|l l|l|l||l|l l|l|l|}
\hline
 Name     &\multicolumn{2}{c|}{ symbols} &mass&width   &  
Name      &\multicolumn{2}{c|}{ symbols} &mass&width     \\
\hline
chargino 1     &\~{}1+&\~{}1-&MC1  &wC1  & mu-sneutrino &\~{}nm&      &MSnm &wSnm \\ 
chargino 2     &\~{}2+&\~{}2-&MC2  &wC2  & tau-sneutrino&\~{}nl&      &MSnl &wSnl \\
neutralino 1   &\~{}o1&      &MNE1 &wNE1 & u-squark L   &\~{}uL&\~{}UL&MSuL &wSuL \\
neutralino 2   &\~{}o2&      &MNE2 &wNE2 & u-squark R   &\~{}uR&\~{}UR&MSuR &wSuR \\
neutralino 3   &\~{}o3&      &MNE3 &wNE3 & c-squark L   &\~{}cL&\~{}CL&MScL &wScL \\
neutralino 4   &\~{}o4&      &MNE4 &wNE4 & c-squark R   &\~{}cR&\~{}CR&MScR &wScR \\
gluino         &\~{}g &      &MSG  &wSG  & t-squark 1   &\~{}t1&\~{}T1&MSt1 &wSt1 \\
selectron L&\~{}eL&\~{}EL&MSeL &wSeL & t-squark 2   &\~{}t2&\~{}T2&MSt2 &wSt2 \\
selectron R&\~{}eR&\~{}ER&MSeR &wSeR & d-squark L   &\~{}dL&\~{}DL&MSdL &wSdL \\
smuon L        &\~{}mL&\~{}ML&MSmL &wSmL & d-squark R   &\~{}dR&\~{}DR&MSdR &wSdR \\
smuon R        &\~{}mR&\~{}MR&MSmR &wSmR & s-squark L   &\~{}sL&\~{}SL&MSsL &wSsL \\
stau 1         &\~{}l1&\~{}L1&MSl1 &wSl1 & s-squark R   &\~{}sR&\~{}SR&MSsR &wSsR \\
stau 2         &\~{}l2&\~{}L2&MSl2 &wSl2 & b-squark 1   &\~{}b1&\~{}B1&MSb1 &wSb1 \\
e-sneutrino    &\~{}ne&      &MSne &wSne & b-squark 2   &\~{}b2&\~{}B2&MSb2 &wSb2 \\
\hline
\end{tabular}
\end{table}

\begin{table*}[htb]
\caption{\label{SMP}Designations for the Standard Model particles}
\vspace{.5cm}
\begin{tabular}{|l|l l|l|l||l|l l|l|l|}
\hline
Name &\multicolumn{2}{c|}{ symbols}  & Mass&Width& Name & \multicolumn{2}{c|}{symbols}  &
Mass&Width \\
\hline
photon       &A  &   & 0 &0& tau-neutrino &nl &Nl & 0 &0  \\
Z boson      &Z  &   & MZ&wZ& tau-lepton   &l  &L  & Ml&0  \\
W boson      &W+ &W- & MW&wW& s-quark      &s  &S  & 0 &0  \\
gluon        &G  &   & 0 &0& c-quark      &c  &C  & 0 & 0 \\
electron     &e  &E  & 0 &0& u-quark      &u  &U  & 0 & 0 \\
muon         &m  &M  & 0 &0& d-quark      &d  &D  & 0 & 0 \\
e-neutrino   &ne &Ne & 0 &0& t-quark      &t  &T  & Mt&wt  \\
mu-neutrino  &nm &Nm & 0 &0& b-quark      &b  &B  & Mb& 0 \\
\hline
\end{tabular}

\end{table*}

\section{Work with the \micro~ package.}

\subsection{Installation and link with RGE packages.}
{\label{INSTALL}
\micro~ can be obtained at\\
\hspace*{2cm}{\tt http://wwwlapp.in2p3.fr/lapth/micromegas}

\noi The name of the file downlaoded should be {\tt
micromegas\_1.3.0.tar.gz}. After unpacking the file, the root
directory of the package, \verb|micromegas_1.3.0|,  will be created.
This directory contains the \verb|micro_make| file, some sample main programs,
a directory for the source code, a directory for \calchep~ interactive sessions and a directory containing  data files.
To compile, type either

\verb|./micro_make|

This command is a Unix script, which detects the operating system and its version, sets
the corresponding compiler options, and  compiles the code.
Being launched  without arguments, \verb|micro_make|  compiles
only auxiliary libraries needed for relic density evaluation.
Otherwise, the  first argument is treated as a {\tt C} or {\tt
Fortran} main program which should be compiled and linked with
these libraries. The  executable file created has  the same  name
as the main program  without  the \verb|.c|/\verb|.f| extension.

It is interesting to investigate the relic density in the
framework of some scenario of supersymmetry breaking. We rely on the public
codes that evaluate the
supersymmetric  spectrum in the context of models defined
at the GUT scale such as the mSUGRA model. One of these packages,
\suspect~\cite{SUSPECT}, is included into the \micro~ package. We
also support an interface with  
\softsusy~\cite{SOFTSUSY}, \spheno~\cite{SPHENO} and
\isajet~\cite{ISAJET}.

To use \isajet,  the corresponding library should be attached to the
code. It can be done via the variable \verb|EXTLIB| to be defined  
in  the \verb|micro_make| file. For example,  to use \isajet~  located in
the \verb|~/isajet769| directory the definition should be\\
\hspace*{2cm}\verb|EXTLIB="$HOME/isajet769/libisajet.a"|\\
 If \verb|mathlib| from CERNLIB is not included in
libisajet.a  it should be specified in EXTLIB, for example \\
\hspace*{2cm}\verb|EXTLIB="$HOME/isajet769/libisajet.a -L/cern/pro/lib -lmathlib"|

The interface with   \softsusy~ and \spheno~   is realized  in the
framework of the {\it SUSY Les Houches accord}\cite{Skands}
by direct execution  of the corresponding programs. In both cases,
the user has to define in the \verb|micro_make| file, the
variables \softsusy~ or \verb|SPHENO|  which identifies the directory where
the corresponding
executable file is located. For example, \\
\hspace*{2cm}\verb|SOFTSUSY=$HOME/softsusy_1.8|\\
or \\
\hspace*{2cm}\verb|SPHENO=$HOME/SPheno2.2.0|

To install the package, one needs initially  about 20MB of disk space. As
the program generates libraries for annihilation processes only at
the time they are required, the total disk space necessary can
double after running the program for different
models as described in the next section.

\subsection{ Dynamic generation of matrix elements and their loading.}
\label{DL} In order to take into account all possible processes of
annihilation of superparticles into SM particles, we need
matrix elements for about 2800 different subprocesses. However, for a given
set of parameters,  usually only a few processes  contribute, other
subprocesses  are suppressed by the Boltzmann factor.

The \micro~ package just after compilation  does not contain the code for
matrix elements. They are generated and linked in runtime  when
needed. To generate the matrix elements we use the CalcHEP program
\cite{CalcHEP}  in  {\it batch} mode \cite{Batch}.  The
compiled matrix elements are stored as {\it shared}
libraries in the subdirectory \\
\hspace*{2cm}\verb|sources/2-2/|\\
The name of the library created corresponds to the names of initial superparticles. Say,
the library containing $\neuto\neuto$ annihilation processes is
\verb|omglib_o1_o1.so|.

On the first few calls,   \micro~  works slowly because it
compiles matrix elements. After being compiled once,  the code for
matrix elements  is stored on the disk and is  accessible for all
subsequent calls. Each process is generated and compiled only
once.

In case several jobs are submitted simultaneously, a problem
occurs  when CalcHEP receives a new request to generate a  matrix
element when it has not completed the previous one. 
We delay the operation of the second program.  The warning that
CalcHEP  is busy  signals the presence of  a {\tt LOCK} file in the directory\\
\hspace*{2cm}\verb|sources/work/tmp|\\
If for some reason this file is not removed after the CalcHEP session,
the user should remove it.

The executable file generated by \verb|micro_make| can be moved and executed
in other directories. However it will  always use and update the matrix elements
stored in\\ \verb|micromegas_1.3.0/sources/2-2|

\subsection{ Linking with other codes and  including  \micro~  into
other packages.}

One can  easily  add  other libraries to the \micro~ package
similarly to the  implementation of {\tt Isajet} described  in
Section~\ref{INSTALL}. One needs to  pass the library name to
the linker via the  EXTLIB variable defined in \verb|micro_make|,
 by  specifying the complete path to the library.
One can  include  the \micro~ package into other C, C++, or
Fortran projects.  The function prototypes for C and C++ projects are
stored in  the  \verb|sources/micromegas.h| file. All the routines
of our package as well as \suspect~ and \feynhiggs~
routines  are stored in  \\
\hspace*{2cm}{\tt sources/micromegas.a} \\
which in turn needs the functions of\\
\hspace*{2cm}{\tt sources/decay2.a} \\
to calculate the widths. 
 The user must  pass to the linker the library that
supports  dynamic loading. The name of this library   depends on
the Unix platform. One can find this name in the \verb|micro_make|
file, it is assigned to the  \verb|LDDL| environment variable.

To attach \micro~ to a C or C++ project, the user 
should make sure that  the library of  Fortran functions are also
 passed to the linker. In the \verb|micro_make| file this library
is described by the  \verb|LDF| variable.

\subsection{Running \micronew: examples.}

\label{examples}
The directory \verb|micromegas_1.3.0| contains several examples of {\it main}
programs. The files \verb|sugomg.c| and \verb|sugomg_f.f| are {\it main}
programs for the evaluation of the relic density in the \verb|mSUGRA| scenario.\\
\hspace*{2cm}\verb|./micro_make sugomg.c|\\
generates the executable  \verb|sugomg| which needs 5 parameters\\
\hspace*{2cm}\verb|./sugomg <m0> <mhf> <a0> <tb> <sgn>|\\
The \verb|sugomg| executable  also understands three additional input parameters
as $\mt$, $\mb(\mb)$, $\alpha_s(\mz)$.
The output contains  the SUSY and Higgs mass spectrum, the value of the  relic density, the relative
contributions of different  processes to $1/\Omega$ as well as
the constraints mentioned in Section~\ref{constrains}.
 The list of
necessary parameters are written on the screen when
 \verb|sugomg| is called without specifying  parameters.\\
\hspace*{2cm}\verb|./micro_make sugomg_f.f|\\
compiles the corresponding Fortran code. In this case the input
parameters are requested  after launching the program:
\begin{verbatim}
      > ./sugomg_f
      Enter  m0   mhf a0   tb  sgn
      >
\end{verbatim}
By default these programs call \suspect~ for solving the  RGE
equations. One can easily change the RGE code
by replacing the \verb|suspectSUGRA| call by the appropriate one
 in \verb|sugomg.c| or  \verb|sugomg_f.f|. 

The program \verb|s_cycle.c| performs the calculation  over
 10   \verb|mSUGRA| test points \cite{benchmark}. 
 Results for these points for
all RGE programs mentioned in our paper are presented in the file
\verb|data/s_cycle.res|.

Finally the \verb|omg.c| and \verb|omg_f.f| programs 
evaluated  the relic density in the case of the unconstrained MSSM.
The input  parameters are read from a  text file written in the
format of the \verb|ewsbInitFile| routine. In the  C-version the file should be passed as a parameter, for example\\
 \verb|     ./omg data/data03|\\
If several sets of parameters are passed to the program, the calculation
will be done in a cycle. The Fortran  version also works in a cycle,
waiting for  a file name as input and finishes  after an empty line
input.

The directory \verb|data| contains 22 \verb|"data*"| test input files
for this routine. 
These parameter sets were
chosen to check the  program in special difficult  cases   where either  strong
co-annihilation and/or Higgs pole contribute  significantly in
relic density evaluation.  Results of relic density calculation
for all these 22 test points using the option when all masses are evaluated at tree-level are stored in file \verb|data\omg.res|.

\subsection{CalcHEP interactive session.}
The CalcHEP \cite{CalcHEP} program  used for matrix element generation is included in the
\micro~\\ package. The user can calculate interactively various
cross sections  both in the general MSSM and in SUGRA models.  To realize this
option the user has to move to the  \verb|calchep| subdirectory and launch \\
\hspace*{2cm}\verb|./calchep|

The implementation of the
MSSM and SUGRA models in \calchep~ is identical to the one in \micro~ described in previous sections.
 There are  two auxiliary parameters, \verb|LCOn| and \verb|dMbOn|
which switch ON/OFF loop corrections to the MSSM particle spectrum and SUSY-QCD
correction to $h,H,A \ra b \bar{b}$ decays respectively.  If \verb|LCOn>0| or
\verb|dMbOn>0|  the corresponding correction is taken into account.

The list of parameters contains also the scale parameter \verb|Q| which should
be set depending of the scale of the process under consideration. This
parameter contributes to the running of $\alpha_s$ and to the  running masses of \verb|t| and
\verb|b| quarks. Here we use the standard $\overline{MS}$  formulae without
including the higher order QCD corrections
\footnote{These corrections can be simulated by decreasing of
scale $Q$. 
}   presented in  Section~\ref{Higgs_decay}.

For the  SUGRA model,  all four RGE packages   
presented in \micro~ can be used, \suspect~ is defined  by default.
External RGE packages  are available for CalcHEP if they were already properly installed  in the  \micro~ package as described in
section~\ref{INSTALL}.
To include another RGE package one has to edit the model in \calchep~ (in the {\it Edit model} menu). 
The   \verb|suspectSUGRA| call should be commented in the {\it Constraints}
menu while the line corresponding to the call for another routine  should be 
uncommented.  The symbol for  comment  is \%.
In the {\it Edit model} menu one can also defined the
non-universal SUGRA model.
By default,  mSUGRA boundary conditions are  implemented. 
To modify this,  first comment the lines  in  the {\it Constraints}
menu  which express the GUT scale parameters in Table~3 in terms of the mSUGRA parameters. The corresponding  non-universal parameters should then be introduced as new variables in the {\it Variables} menu.

In this realization of MSSM/SUGRA all widths of super-partners are
evaluated automatically at tree-level including all  \verb|1->2| 
decay modes generated in the model.
The relic  density and other constrains mentioned in section~\ref{constrains}
are included in the list of {\it Constrains} and automatically attached to
CalcHEP numerical sessions. 

In CalcHEP numerical sessions for \verb|2->2| processes we provide an
option to   construct a plot for the $v\cdot\sigma$ dependence on the incoming
momentum. This option is found under the  {\it Simpson} menu
function.

\subsection{Sample output file}

Running \micronew~ with the default values of the standard parameters and
choosing  the \suspect~ RGE package with the mSUGRA input parameters 	

\verb|sugomg 107 600 0 5 1|

\noi
will produce the following output:

{\small
\begin{verbatim}
Higgs masses and widths
h   : Mh    =  116.0 (wh    =2.5E-03)
H   : MHH   =  899.2 (wHh   =1.9E+00)
H3  : MH3   =  898.5 (wH3   =2.2E+00)
H+  : MHc   =  902.0 (wHc   =2.3E+00)

Masses of SuperParticles:
~o1 : MNE1  =   249.1 || ~l1 : MSl1  =   254.2 || ~eR : MSeR  =   256.0 
~mR : MSmR  =   256.0 || ~nl : MSnl  =   413.1 || ~ne : MSne  =   413.4 
~nm : MSnm  =   413.4 || ~eL : MSeL  =   420.2 || ~mL : MSmL  =   420.2 
~l2 : MSl2  =   420.4 || ~1+ : MC1   =   468.3 || ~o2 : MNE2  =   468.5 
~o3 : MNE3  =   780.0 || ~2+ : MC2   =   793.2 || ~o4 : MNE4  =   794.3 
~t1 : MSt1  =   946.7 || ~b1 : MSb1  =  1153.1 || ~b2 : MSb2  =  1187.8 
~dR : MSdR  =  1188.4 || ~sR : MSsR  =  1188.4 || ~t2 : MSt2  =  1190.6 
~uR : MSuR  =  1194.8 || ~cR : MScR  =  1194.8 || ~uL : MSuL  =  1248.2 
~cL : MScL  =  1248.2 || ~dL : MSdL  =  1250.5 || ~sL : MSsL  =  1250.5 
~g  : MSG   =  1358.1 || 
Xf=2.67e+01 Omega=8.87e-02

Channels which contribute to 1/(omega) more than 1%.
Relative contrubutions in % are displyed
  1% ~o1 ~o1 -> l L 
  3% ~o1 ~l1 -> Z l 
 12% ~o1 ~l1 -> A l 
  2% ~o1 ~eR -> Z e 
  8% ~o1 ~eR -> A e 
  2% ~o1 ~mR -> Z m 
  8% ~o1 ~mR -> A m 
 11% ~l1 ~l1 -> l l 
  2% ~l1 ~L1 -> A Z 
  3% ~l1 ~L1 -> A A 
  8% ~eR ~l1 -> e l 
  6% ~eR ~eR -> e e 
  1% ~eR ~ER -> A Z 
  2% ~eR ~ER -> A A 
  6% ~eR ~mR -> e m 
  8% ~mR ~l1 -> m l 
  6% ~mR ~mR -> m m 
  1% ~mR ~MR -> A Z 
  2% ~mR ~MR -> A A 
deltartho=9.11E-06
gmuon=3.12E-10
bsgnlo=3.85E-04
bsmumu=3.13E-09
MassLimits OK

\end{verbatim}
}

Under the same conditions and for the same set of  parameters, running
the cross-section and branching ratios routines 

\verb|cs_br|

\noi
will produce the following output:

{\small
\begin{verbatim}
Example of some cross sections and widths calculation
  for mSUGRA point m0=107.0,mhf=600.0,a0=0.0,tb=5.0

 Z partial widths
  b   B - 3.684E-01 GeV
  d   D - 3.703E-01 GeV
  u   U - 2.873E-01 GeV
  c   C - 2.873E-01 GeV
  s   S - 3.703E-01 GeV
  l   L - 8.378E-02 GeV
 nl  Nl - 1.670E-01 GeV
 nm  Nm - 1.670E-01 GeV
 ne  Ne - 1.670E-01 GeV
  m   M - 8.397E-02 GeV
  e   E - 8.397E-02 GeV
Total 2.436E+00 GeV

 h partial widths
  b   B - 2.460E-03 GeV
  l   L - 2.552E-04 GeV
Total 2.716E-03 GeV

Cross sections at Pcm=500.0 GeV
e,E->~1+,~1- 
e,E->~1+(468),~1-(468)  is 7.135E-03 pb
e,E->~o1,~o2 
e,E->~o1(249),~o2(468)  is 1.130E-02 pb

\end{verbatim}
}

\section{Results}

We have compared the results obtained with \micronew~ and those obtained with \darksusy4.0 for 10 benchmarks mSUGRA points\cite{benchmark}. 
For this check, we have used \isajet7.69,  $m_t^{pole}=174.3$GeV,
$\alpha_s(\mz)=.1172, \mbmb=4.23$GeV.
The latter is only relevant for the calculation of the Higgs widths.

As seen in Table~\ref{darksusy}, the two programs agree at the 3\% level except 
at large $\tan\beta$.  This discrepancy  is due to a difference
in the width of the pseudoscalar.
We recover good agreement with  \darksusy~ (below 3\%) if
we substitute their value for the pseudoscalar width. 

\begin{table*}[htb]
\caption{\label{darksusy} 
Comparison between \micronew~ and \darksusy4.0}
\vspace{.5cm}
\begin{tabular}{|c|ccccc|l|l|}
\hline
name  &$\m0$&$\mhf$  & $A_0$ & $\tan\beta$ & $sgn(\mu)$    &\micronew      & \darksusy4.0\\
\hline
A& 107& 600&0& 5& 1& 0.0944& 0.0929\\
B& 57& 250&0& 10& 1& 0.124& 0.121\\
C& 80& 400&0& 10& -1& 0.117& 0.115\\
D& 101& 525&0& 20& 1& 0.0876& 0.0864\\
G& 113& 375&0& 20& 1& 0.133& 0.129\\
H& 244& 935&0& 20& 1& 0.166& 0.163\\
I& 181& 350&0& 35& 1& 0.142& 0.132\\
J& 299& 750&0& 35& 1& 0.102& 0.0975\\
K& 1001& 1300&0& 46& -1& 0.0893& 0.0870\\
L& 303& 450&0& 47& 1& 0.114& 0.0982\\
\hline
\end{tabular}
\end{table*}

\section{Conclusion}

\micronew~ solves with an accuracy at the percent level
the evolution equation for the density of supersymmetric particles and calculates  the relic density  of dark matter.
All possible channels for annihilation and coannihilations are included
and all matrix elements are calculated exactly in an improved tree-level
approximation that uses 
pole masses and loop-corrected mixing matrices for supersymmetric particles.
Loop corrections to the masses of Higgs particles and to the partial widths of the Higgs (QCD and SUSY) are implemented.
These higher-order corrections are essential 
since  the annihilation cross-section can be  very sensitive to the
mass of the particles that contribute to the various annihilation processes, 
in particular  near a resonance or in regions of parameter space where coannihilations occur. Furthermore, both these processes are often the dominant ones in physically interesting supersymmetric models,
that is in models where the relic density is below the WMAP upper limit.  

The relic density can be calculated starting from a set of MSSM parameters defined at the weak scale or at the GUT scale. We provide an interface to the four major codes that calculate the supersymmetric spectrum using renormalization group equations.
Within the context of the mSUGRA model, there are still large uncertainties in the computation of the supersymmetric spectrum \cite{ben}, this of course will have a strong impact on the prediction for the relic density \cite{houches_compar}.
An accurate prediction of the relic density within SUGRA models then 
presupposes a precise knowledge of the supersymmetric spectrum.

New features of the package also include  the computation of cross-sections and decay widths for any process in the MSSM with two-body final states as well as
an  improved NLO calculation of the $\bsgamma$  branching ratio and a  new routine for the $\bsmu$ decay rate.

\section{Acknowledgements}

We thank A.~Cottrant for providing the code for the $(g-2)_\mu$.
We have also benefitted from discussions with  B.~Allanach, A.~Belyaev, A.~Djouadi, J.~L.~Kneur and W.~Porod on the RGE codes.
 We would like to thank M. Gomez for testing parts of  our code.
 We also thank P.~Gambino 
 for discussion and for some clarification regarding
\cite{DGG}, in particular for confirming our implementation of the
large $\tgb$ effects in SUSY.
This work was supported in part by the PICS-397, {\it Calcul en physique des particules},by GDRI-ACPP of CNRS and by grants from the
Russian Federal Agency for Science, NS-1685.2003.2  and RFBR 04-02-17448.

\appendix
\section*{Appendix}

\section{List of functions}


\subsection{ \micro~ functions in C.}

{\small
\begin{verbatim}
int assignVal(char * name, double val)
void assignValW(char * name, double val)
int readLesH(char *fname)
int ewsbInitFile(char * fname,int LC)}
int ewsbMSSM(tb,MG1,MG2,MG3,Am,Al,At,Ab,MH3,mu,Ml1,Ml2,Ml3,Mr1,Mr2,Mr3,Mq1,Mq2,Mq3,
    Mu1,Mu2,Mu3,Md1,Md2,Md3,LC); int LC; all other parameters are 'double'
int xxxxxSUGRA(tb,MG1,MG2,MG3,Al,At,Ab,sgn,MHu,MHd,Ml1,Ml3,Mr1,Mr3,Mq1,Mq3,
                  Mu1,Mu3,Md1,Md3)
\end{verbatim}
\noi
All parameters are `double'. \verb|'xxxx'| is \verb|'suspect','isajet','softSusy',|or \verb|'spheno'|
\begin{verbatim}	  
int calcDep(int dMbOn)
int findVal(char * name, double * val)
double findValW(char * name)
void printVar(FILE *f, int N)
void printMasses(FILE * f, int sort)
char * lsp(void)
double lspmass_()
double darkOmega(double *Xf,int fast, double Beps)
double darkOmegaFO(double *Xf,int fast, double Beps)
double printChannels(double Xf, double cut, double Beps, int prcnt, FILE *f )
double deltarho_(void)
double bsgnlo_(void)
double bsmumu_(void)
double gmuon_(void)
int    masslimits_(void)
double MbRun(double Q)
double MtRun(double Q)
double MbEff(double Q)
double MtEff(double Q)
double deltaMb(void)
double decay2(char*pIn, int k, char*pOut1, char*pOut2)
void* newProcess(char* procName, char*libName)
int infor22(void*address,int nsub, char*pIn1,char*pIn2,char*pOut1,char*pOut2,
             double*m1,double*m2,double*m3,double*m4)
double cs22(void*address, int nsub, double Pcm, double c1, double c2, int*err)
double  annihilation(double v, int k, char * pOut1, char pOut2)
\end{verbatim}
}

\noi
{\bf \micro functions in Fortran.}

{\small
\begin{verbatim}
  INTEGER FUNCTION assignVal(name,val)
  SUBROUTINE       assignValW(name,val)
  INTEGER FUNCTION readLesH(fname)
  INTEGER FUNCTION ewsbInitFile(fname,LC)
  INTEGER FUNCTION ewsbMSSM(tb,MG1,MG2,MG3,Am,Al,At,Ab,MH3,mu,Ml1,Ml2,Ml3,
         Mr1,Mr2,Mr3,Mq1,Mq2,Mq3, Mu1,Mu2,Mu3,Md1,Md2,Md3,LC)
  INTEGER FUNCTION xxxxSUGRA(tb,MG1,MG2,MG3,Al,At,Ab,sgn,MHu,MHd,
         Ml1,Ml3,Mr1,Mr3,Mq1,Mq3,Mu1,Mu3,Md1,Md3)
\end{verbatim}
\noi
All parameters are `double'. \verb|'xxxx'| is \verb|'suspect','isajet','softSusy',|or \verb|'spheno'|
\begin{verbatim}	  
  INTEGER FUNCTION calcDep(dMbOn)
  INTEGER FUNCTION findVal(name, val)
  REAL*8  FUNCTION findValW(name)
  SUBROUTINE       printVar(n)
  SUBROUTINE       printMasses(sort) 
  SUBROUTINE       LSP(name)
  REAL*8  FUNCTION lspMass()
  REAL*8  FUNCTION darkOmega(Xf,fast,Beps)
  REAL*8  FUNCTION darkOmegaFO(Xf,fast,Beps)
  REAL*8  FUNCTION printChannels(Xf,cut,Beps,prcnt)
  REAL*8  FUNCTION deltarho()
  REAL*8  FUNCTION bsgnlo()
  REAL*8  FUNCTION bsmumu()
  REAL*8  FUNCTION gmuon()
  INTEGER FUNCTION MassLimits()
  REAL*8  FUNCTION MbRun(Q)
  REAL*8  FUNCTION MtRun(Q)
  REAL*8  FUNCTION MbEff(Q)
  REAL*8  FUNCTION MtEff(Q)
  REAL*8  FUNCTION deltaMb()
  REAL*8  FUNCTION decay2(pIn, k, pOut1,pOut2)
  SUBROUTINE newProcess(procName,libName,address)
  INTEGER FUNCTION infor22(address,nsub, pIn1,pIn2,pOut1,pOut2,m1,m2,m3,m4)
  REAL*8  FUNCTION cs22(address, nsub, Pcm, c1, c2 , ERR)
  REAL*8  FUNCTION annihilation(v,k, pOut,pOut2)
\end{verbatim}}

\noi
The types of the  parameters are:
{\begin{verbatim}
  CHARACTER  pIn*(*),pIn1*(*),pIn2*(*),pOut1*(*),pOut2*(*),
 >  name*(*),fname*(*),procName,*(*),libName(*,*)  
  REAL*8   val,Xf,Beps,cut,Pcm,c1,c2,v,Q,m1,m2,m3,m4
  REAL*8   tb,MG1,MG2,MG3,Am,Al,At,Ab,MH3,mu,Ml1,Ml2,Ml3,
 >       Mr1,Mr2,Mr3,Mq1,Mq2,Mq3, Mu1,Mu2,Mu3,Md1,Md2,Md3
  INTEGER    n,k,sort,prcnt,ERR,LC,dmbOn,fast, address[2]
\end{verbatim}
}

\section{Implementation of {\boldmath $B(B\ra s\gamma)$}
in
\bf \tt micrOMEGAs}

 The calculation  for $B(B\ra s\gamma)$ in the MSSM is quite involved
and requires that one goes beyond one-loop. Most of what is
described below, as implemented in {\tt micrOMEGAs}, is in fact
just a, unified, compendium of different contributions that have
appeared in the literature.
 There is no claim of originality, most
expressions are taken verbatim. However care has been taken in
carefully checking all formulae that have appeared in the
literature. This has helped, for example, identify a few misprints
and typos and allowed to generalise some results. By giving the
detail of the implementation, it is possible to easily modify this
routine of the {\tt micrOMEGAs} code in order to include future
new contributions both to the SM and the MSSM.
Note that we redefine in this routine many parameters
used in \micronew, for example the running quark masses, this
routine can then be used as a stand-alone routine.

\subsection{General set-up: From $M_W$ to $\mu_b$, QED corrections}
Our implementation of the Standard Model contribution follows the
work of  Kagan and Neubert \cite{KN} very closely. We however
include the effect of a running $c$ quark mass heuristically so
that our results take into account the latest calculations of
Gambino and Misiak\cite{Gambino-Misiak} who advocate the use of
the ${\overline{MS}}$ charm mass, $m_c(m_b)$. The ({\em relevant})
operator basis is
\begin{eqnarray}
   O_2 &=& \bar s_L\gamma_\mu c_L\bar c_L\gamma^\mu b_L \,, \nonumber\\
   O_7 &=& \frac{e\,m_b}{16\pi^2}\,\bar s_L\sigma_{\mu\nu}
    F^{\mu\nu} b_R \,, \nonumber\\
   O_8 &=& \frac{g_s m_b}{16\pi^2}\,\bar s_L\sigma_{\mu\nu}
    G_a^{\mu\nu} t_a b_R \,.
\end{eqnarray}
 which defines
\begin{equation}
   H_{\rm eff} = -\frac{4 G_F}{\sqrt2}\,V_{ts}^* V_{tb}
   \sum_i C_i(\mu_b) O_i(\mu_b) \,.
   \label{hamiltonian}
\end{equation}

The renormalisation scale $\mu_b$ in (Eq.~\ref{hamiltonian}) is of
order $m_b$ and is usually let to vary in the range $(m_b/2,2
m_b)$. The default value in the code is $m_b$. Varying $\mu_b$ is
one measure of the theoretical error. The branching fraction
writes

\begin{equation}
   {\rm B}(B\to X_s\gamma)
   =\frac{6\alpha}{\pi f(z_0)}\,\left| \frac{V_{ts}^* V_{tb}}{V_{cb}}
    \right|^2 K_{\rm NLO}(\delta)\;\times\; {\rm B}(B\to
    X_c\,e\,\bar\nu)\;.
\label{numeric}
\end{equation}

By default we take
\beqn
{\rm B}(B\to X_c\,e\,\bar\nu)=0.1045
\eeqn

The kinematical function, $f(z_0)$, is defined as
\beqn
f(z_0)=1-8z_0+8z_0^3-z_0^4-12z_0^2\ln z_0\approx 0.542-2.23(\sqrt
z_0-0.29)
\eeqn
with $z_0=(m_c/m_b)^2$ defined in terms of the {\em pole} masses,
giving a value in the range,$\sqrt z_0=0.29\pm 0.02$. For the
radiative photon we take  $\alpha=1/137.036$.

The factor $K_{\rm NLO}(\delta)$ involves the  photon energy
cut-off parameter $\delta$ that shows up at the NLO. In {\micro} this value is set  to $\delta=0.9$ as is generally
assumed in order to describe the ``total" (fake) branching ratio.
With the formulae given below the code can be modified in a very
straightforward way to take into account the full  $\delta$
dependence. $K_{\rm NLO}(\delta)$ is decomposed in terms of the
Wilson coefficients with leading (LO) and next-to-leading (NLO) contributions
as

\begin{eqnarray}
   K_{\rm NLO}(\delta) &=& \sum_{ \stackrel{i,j=2,7,8}{i\le j} }
    k_{ij}(\delta,\mu_b)\,\mbox{Re}\!\left[ C_i^{(0)}(\mu_b)\,
    C_j^{(0)*}(\mu_b) \right]
    + S(\delta)\,\frac{\alpha_s(\mu_b)}{2\pi}\,
    \mbox{Re}\!\left[ C_7^{(1)}(\mu_b)\,C_7^{(0)*}(\mu_b) \right]
    \nonumber\\
   &&\mbox{}+ S(\delta)\,\frac{\alpha}{\alpha_s(\mu_b)} \bigg(
    2\,\mbox{Re}\!\left[ C_7^{({\rm em})}(\mu_b)\,C_7^{(0)*}(\mu_b)
    \right] - k_{\rm SL}^{({\rm em})}(\mu_b)\,|C_7^{(0)}(\mu_b)|^2
    \bigg) \,,
\label{KNLO}
\end{eqnarray}

and
\begin{equation}
   C_i(\mu_b) = C_i^{(0)}(\mu_b) + \frac{\alpha_s(\mu_b)}{4\pi}\,
   C_i^{(1)}(\mu_b) + \frac{\alpha}{\alpha_s(\mu_b)}\,
   C_i^{({\rm em})}(\mu_b) + \dots \,.
\label{Ciexp}
\end{equation}

The leading-order coefficients at the low scale $\mu_b$ of order
$m_b$  are given by
\begin{eqnarray}
   C_2^{(0)}(\mu_b) &=& \frac 12 \left( \eta^{-\frac{12}{23}}
    + \eta^\frac{6}{23} \right) \,, \nonumber\\
   C_7^{(0)}(\mu_b) &=& \eta^\frac{16}{23}\,C_7^{(0)}(m_W)
    + \frac 83 \left( \eta^\frac{14}{23} - \eta^\frac{16}{23} \right)
    C_8^{(0)}(m_W) + \sum_{i=1}^8\,h_i\,\eta^{a_i} \,, \nonumber\\
   C_8^{(0)}(\mu_b) &=& \eta^\frac{14}{23}\,\left( C_8^{(0)}(m_W)
   + \frac{313063}{363036} \right)
    + \sum_{i=1}^4\, h_i^{(8)}\,\eta^{b_i} \,,
\label{evolvemwmb}
\end{eqnarray}
where $\eta=\alpha_s(m_W)/\alpha_s(\mu_b)$, and $h_i$, $
h_i^{(8)}$ and $a_i$ are known {\em numerical} coefficients
\cite{CMM}.

\beqn
h_i&=&(\frac{626126}{272277},-\frac{56281}{51730},-\frac{3}{7},-\frac{1}{14},-0.6494,-0.0380,-0.0186,-0.0057)
\nonumber \\
a_i&=&(\frac{14}{23},\frac{16}{23},\frac{6}{23},-\frac{12}{23},0.4086,-0.4230,-0.8994,0.1456)
\nonumber \\
h_i^{(8)}&=&(-0.9135,0.0873,-0.0571,0.0209)\nonumber \\
b_i&=&(0.4086,-0.4230,-0.8994,0.1456)
\eeqn

For the running of $\alpha_s$ between the scale $M_W$ and $\mu_b$
we use the SM running with 5 flavours which, to a very good
precision, can be implemented as:
\beqn
\alpha_s(\mu)=\frac{\alpha_s(M_Z)}{v_s(\mu)}
\left(1-\frac{116}{23} \frac{\alpha_s(M_Z)}{4 \pi} \frac{\ln
(v_s(\mu))}{v_s(\mu)} \right) \;\;\;\;v_s(\mu)=1-\frac{23}{3}
\frac{\alpha_s(M_Z)}{2 \pi} \ln(M_Z/\mu) \nonumber \\
\eeqn

The value of $\alpha_s(M_Z)$ is read in by the main code {\tt
micrOMEGAs}. For the numerical values that we quote in this note,
we take the default $\alpha_s(M_Z)=0.1185$.

The next-to-leading Wilson coefficient at $\mu_b$,
$C^{(1)}_7(\mu_b)$ is implemented according to \cite{CMM},
\beqn
    \label{c7eff1} C^{(1)eff}_7(\mu_b) &=& \eta^{\frac{39}{23}}
C^{(1)eff}_7(M_W) + \frac{8}{3} \left( \eta^{\frac{37}{23}} -
\eta^{\frac{39}{23}} \right) C^{(1)eff}_8(M_W) \nonumber \\ &&
+\left( \frac{297664}{14283}
\eta^{\frac{16}{23}}-\frac{7164416}{357075} \eta^{\frac{14}{23}}
       +\frac{256868}{14283} \eta^{\frac{37}{23}} -\frac{6698884}{357075} \eta^{\frac{39}{23}} \right) C_8^{(0)}(M_W)
\nonumber \\ && +\frac{37208}{4761} \left( \eta^{\frac{39}{23}} -
\eta^{\frac{16}{23}} \right) C_7^{(0)}(M_W) + \sum_{i=1}^8 (e_i
\eta E(x) + f_i + g_i \eta) \eta^{a_i},
\eeqn

\beqn
\begin{array}{ccccccccc}
\vspace{0.2cm} e_i = ( &\frac{4661194}{816831},&
-\frac{8516}{2217}, & 0,    &     0,
        &     -1.9043,       &       -0.1008,     &  0.1216,  &  0.0183    )\\
\vspace{0.2cm} f_i = ( &    -17.3023,       &        8.5027,     &
4.5508,  &  0.7519,
        &      2.0040,       &        0.7476,     & -0.5385,  &  0.0914    )\\
\vspace{0.2cm} g_i = ( &     14.8088,       &      -10.8090,     &
-0.8740,  &  0.4218,
        &     -2.9347,       &        0.3971,     &  0.1600,  &  0.0225    )\\
\end{array}
&& \nonumber
\eeqn

and

\beqn E(x) = \frac{x (18 -11 x - x^2)}{12 (1-x)^3} + \frac{x^2 (15 -
16 x + 4 x^2)}{6 (1-x)^4} \ln x-\frac{2}{3} \ln x.
\eeqn

The QED coefficients  $C_7^{({\rm em})}(\mu_b)$ and $k_{\rm
SL}^{({\rm em})}(\mu_b)$ are

The result for $C_7^{({\rm em})}(\mu_b)$ is
\begin{eqnarray}
   C_7^{({\rm em})}(\mu_b) &=&
    \left( \frac{32}{75}\,\eta^{-\frac{9}{23}}
    - \frac{40}{69}\,\eta^{-\frac{7}{23}}
    + \frac{88}{575}\,\eta^{\frac{16}{23}} \right) C_7^{(0)}(m_W)
    \nonumber\\
   &&\mbox{}+
    \left( -\frac{32}{575}\,\eta^{-\frac{9}{23}}
    + \frac{32}{1449}\,\eta^{-\frac{7}{23}}
    + \frac{640}{1449}\,\eta^{\frac{14}{23}}
    - \frac{704}{1725}\,\eta^{\frac{16}{23}} \right) C_8^{(0)}(m_W)
    \nonumber\\
   &&\mbox{}-\frac{190}{8073}\,\eta^{-\frac{35}{23}}
    - \frac{359}{3105}\,\eta^{-\frac{17}{23}}
    + \frac{4276}{121095}\,\eta^{-\frac{12}{23}}
    + \frac{350531}{1009125}\,\eta^{-\frac{9}{23}} \nonumber\\
   &&\mbox{}+ \frac{2}{4347}\,\eta^{-\frac{7}{23}}
    - \frac{5956}{15525}\,\eta^{\frac{6}{23}}
    + \frac{38380}{169533}\,\eta^{\frac{14}{23}}
    - \frac{748}{8625}\,\eta^{\frac{16}{23}} \,.
   \nonumber\\
\label{fancy}
\end{eqnarray}

\begin{equation}
   k_{\rm SL}^{({\rm em})}(\mu_b)
   = \frac{12}{23} \left( \eta^{-1} - 1 \right)
   = \frac{2\alpha_s(\mu_b)}{\pi} \ln\frac{m_W}{\mu_b} \,.
\label{kSL}
\end{equation}

The coefficient functions $k_{ij}(\delta,\mu_b)$ in (\ref{KNLO})
are given by \cite{KN}
\begin{eqnarray}
   k_{77}(\delta,\mu_b) &=& S(\delta) \left\{ 1
    + \frac{\alpha_s(\mu_b)}{2\pi} \left( r_7
    + \gamma_{77}\ln\frac{m_b}{\mu_b} - \frac{16}{3} \right)
    + \left[ \frac{(1-z_0)^4}{f(z_0)} - 1\right]
    \frac{6\lambda_2}{m_b^2} \right\} \nonumber\\
   &&\mbox{}+ \frac{\alpha_s(\mu_b)}{\pi}\,f_{77}(\delta)
    + S(\delta)\,\frac{\alpha_s(\bar\mu_b)}{2\pi}\,\bar\kappa(z) \,,
    \nonumber\\
   k_{27}(\delta,\mu_b) &=& S(\delta) \left[
    \frac{\alpha_s(\mu_b)}{2\pi}
    \left( \mbox{Re}(r_2) + \gamma_{27}\ln\frac{m_b}{\mu_b} \right)
    - \frac{\lambda_2}{9  \;m_b^2 z_0} \right]
    + \frac{\alpha_s(\mu_b)}{\pi}\,f_{27}(\delta) \,, \nonumber\\
   k_{78}(\delta,\mu_b) &=& S(\delta)\,\frac{\alpha_s(\mu_b)}{2\pi}
    \left( \mbox{Re}(r_8) + \gamma_{87}\ln\frac{m_b}{\mu_b} \right)
    + \frac{\alpha_s(\mu_b)}{\pi}\,f_{78}(\delta) \,, \nonumber\\
   k_{ij}(\delta,\mu_b) &=& \frac{\alpha_s(\mu_b)}{\pi}\,
    f_{ij}(\delta) \,;\qquad \{i,j\}=\{2,2\},\,\{8,8\},\,\{2,8\} \,,
\label{kij}
\end{eqnarray}
with the Sudakov factor
\begin{equation}
   S(\delta) = \exp\!\left[ -\frac{2\alpha_s(\mu_b)}{3\pi}
   \left( \ln^2\!\delta + \frac 72\ln\delta \right) \right]
\label{Suda}
\end{equation}
and $\gamma_{77}=\frac{32}{3}$, $\gamma_{27}= \frac{416}{81}$ and
$\gamma_{87}=-\frac{32}{9}$ are entries of the anomalous dimension
matrix. The value of the hadronic parameter is
$\lambda_2=0.12$GeV$^2$.

For these functions we deviate slightly from KN\cite{KN} in the
sense that we define $z_0$ in terms of the pole masses in the
kinematics factor and also $z$ that differs from $z_0$ by the use
of the $\overline{MS}$ running charm mass, $m_c(m_b)$, as
advocated recently by Gambino and Misiak\cite{Gambino-Misiak} in
order to reduce the NNLO uncertainty. We then take 
\beqn
\label{z22}
\sqrt z=0.22\pm.04
\eeqn
 everywhere else in Eq.~\ref{kij}.

The other coefficients are given by

\begin{eqnarray}
\label{eq17}
   r_7 &=& -\frac{10}{3} - \frac{8\pi^2}{9} \,, \qquad
    \mbox{Re}(r_8) = \frac{44}{9} - \frac{8\pi^2}{27} \,, \nonumber\\
   \mbox{Re}(r_2) &\approx& -4.987 + 12.78(\sqrt z-0.22) \,,
   \qquad \bar\kappa(z)\approx 3.672-4.14(\sqrt z-0.22)
\end{eqnarray}

Note that the scale $\bar\mu_b$, in Eq.~\ref{kij}, of relevance in
semileptonic $B$ decays  is in principle different from
the one in the radiative decay.

The real-gluon radiation functions $f_{ij}(\delta)$ can be  coded
for a general photon energy cut-off. They are taken from \cite{KN}
and write as

\begin{eqnarray}
   f_{77}(\delta) &=& \frac{1}{3}\,\bigg[ 10\delta + \delta^2
    - \frac{2\delta^3}{3} + \delta(\delta-4)\ln\delta \bigg] \,,
    \nonumber\\
   f_{88}(\delta) &=& \frac{1}{27}\,\Bigg\{ 4 Li_2(1-\delta)
    - \frac{2\pi^2}{3} + 8\ln(1-\delta) - \delta(2+\delta)\ln\delta
    \nonumber\\
   &&\hspace{0.7cm}\mbox{}+ 7\delta + 3\delta^2 - \frac{2\delta^3}{3}
    - 2 \left[ 2\delta + \delta^2 + 4\ln(1-\delta) \right]
    ln_{bs} \Bigg\} \,, \nonumber\\
ln_{bs}&=&\ln\frac{m_b}{m_s} \;\;\; {\rm we \; take} \;
\frac{m_b}{m_s} \simeq 50 \,, \nonumber\\
   f_{78}(\delta) &=& \frac{8}{9}\,\Bigg[ Li_2(1-\delta)
    - \frac{\pi^2}{6} - \delta\ln\delta + \frac{9\delta}{4}
    - \frac{\delta^2}{4} + \frac{\delta^3}{12} \Bigg] \,, \nonumber\\
   f_{22}(\delta) &=& \frac{16}{27} \int\limits_0^1\!{\rm d}x\,
    (1-x)(1-x_\delta)\,\left|\, \frac{z}{x}\,
    G\!\left(\frac{x}{z}\right) + \frac 12 \,\right|^2 \,, \nonumber\\
   f_{27}(\delta) &=& -3 f_{28}(\delta)
    = - \frac{8z}{9} \int\limits_0^1\!{\rm d}x\,(1-x_\delta)\,
    \mbox{Re}\!\left[ G\!\left(\frac{x}{z}\right) + \frac{x}{2z}
    \right] \,,
\label{fij}
\end{eqnarray}
where $x_\delta=\mbox{max}(x,1-\delta)$, and
\begin{equation}
   G(t) = \left\{ \begin{array}{cl}
    -2\arctan^2\!\sqrt{t/(4-t)} & ~;~t<4 \,, \\[0.1cm]
    2 \left( \ln\Big[(\sqrt{t}+\sqrt{t-4})/2\Big]
    - \displaystyle\frac{i\pi}{2} \right)^2 & ~;~t\ge 4 \,.
   \end{array} \right.
\end{equation}

Since we will specialise to the case $\delta=0.9$, it is more
efficient to quote the corresponding values of $f_{ij}$, and give
approximations to $f_{27},f_{22}$ that we use in the code:

\beqn
f_{77}(0.9) &=&  3.20599\,,
    \nonumber\\
   f_{88}(0.9) &=&  1.31742\,, \nonumber\\
   f_{78}(0.9) &=& 0.387341 \,, \nonumber\\
   f_{22}(0.9) &\approx& 0.107636 \;-\;0.208484 \sqrt{z}\;-\;0.156146 z=0.05421-0.2772 \epsilon_z-0.156146 \epsilon_z^2 \,, \nonumber\\
   f_{27}(0.9) &=& -3 f_{28}(0.9) \approx -0.190805 \;+\;0.948865 \sqrt{z}\;-\;0.787805
   z \, \nonumber\\
   &=&-0.02023+.6020 \epsilon_z-0.7878 \epsilon_z^2\; .\nonumber\\
   \epsilon_z&=&\sqrt{z}-0.22
\label{fijapprox}
\end{eqnarray}

Taking $z=0.22^2$, Eq.\ref{z22}, rather than $z=0.29^2$ mainly affects $K_{27}$
especially through $Re(r_2)$. Note the large coefficient of
$\epsilon_z$ in $Re(r_2)$ in Eq.~\ref{eq17}.

\subsection{Standard Model contribution}
This contribution we take from \cite{CDGG-smh}.
 We first define the functions
\beq
F_7^{(1)}(x)=\frac{x(7-5x-8x^2)}{24(x-1)^3}+\frac{x^2(3x-2)}{4(x-1)^4}\ln
x \label{f71}
\eeq
\beq
F_8^{(1)}(x)=\frac{x(2+5x-x^2)}{8(x-1)^3}-\frac{3x^2}{4(x-1)^4}\ln
x \label{f81}
\eeq
\subsubsection{LO at $M_W$}
We have
\beq
C^{(0){\rm SM}}_{7,8}(\mu_W)=F_{7,8}^{1}(x_{tw})
\eeq
where $x_{tw}$ is defined in terms of the running top mass at the
weak scale, $\mu_W=M_W$.
\beq
x_{tw}=\frac{\overline{m}_t^2(\mu_W )}{M_W^2}.
\eeq
For the  NLO top-quark running mass at the scale $\mu_W$ we follow
\cite{CDGG-smh}
\beqn
\overline{m}_t (\mu_W )&=&\overline{m}_t (m_t) \left[
\frac{\alpha_s (\mu_W)}{\alpha_s (m_t )} \right]^{\frac{12}{23}}
\left[ 1+ \frac{\alpha_s
(m_t)}{4\pi}\frac{\gamma_0^m}{2\beta_0}\left(
\frac{\gamma_1^m}{\gamma_0^m}-\frac{\beta_1}{\beta_0}\right)
\left( \frac{\alpha_s (\mu_W )}{\alpha_s (m_t)}-1\right) \right]
 \nonumber
\\ \overline{m}_t(m_t)&=&m_t\left[ 1-\frac{4}{3}\frac{\alpha_s
(m_t)}{\pi}\right]\;\;,\;\;\overline{m}_t(m_t)^2=m_t^2\left[
1-\frac{8}{3}\frac{\alpha_s (m_t)}{\pi}\right]
\eeqn
$m_t$ is the pole mass which in this note we take as $m_t=174.3\pm
5.1$GeV for comparison with other authors.
\beq
\beta_0=\frac{23}{3}\;,\; \beta_1=\frac{116}{3}\;,\;\gamma_0^m= 8
\;,\;\gamma_1^m=\frac{1012}{9}
\eeq

\subsubsection{NLO SM}

We have
\beqn
C^{(1){\rm SM}}_{7,8}(M_W)=G_{7,8}^{1}(x_{tw})
\eeqn

\beqn
G_7 (x) &=& \frac{-16 x^4 -122 x^3 + 80 x^2 -  8 x}{9 (x-1)^4}
{\rm Li}_2 \left( 1 - \frac{1}{x} \right)
                  +\frac{6 x^4 + 46 x^3 - 28 x^2}{3 (x-1)^5} \ln^2 x
\nonumber \\ &&
                  +\frac{-102 x^5 - 588 x^4 - 2262 x^3 + 3244 x^2 - 1364 x +
208} {81 (x-1)^5} \ln x \nonumber \\ &&
                  +\frac{1646 x^4 + 12205 x^3 - 10740 x^2 + 2509 x - 436}
{486 (x-1)^4} \label{c71eff}
\vspace{0.2cm} \\
G_8(x) &=& \frac{-4 x^4 +40 x^3 + 41 x^2 + x}{6 (x-1)^4} {\rm
Li}_2 \left( 1 - \frac{1}{x} \right)
                  +\frac{ -17 x^3 - 31 x^2}{2 (x-1)^5} \ln^2 x
\nonumber \\ &&
                  +\frac{ -210 x^5 + 1086 x^4 +4893 x^3 + 2857 x^2 - 1994 x
+280} {216 (x-1)^5} \ln x \nonumber \\ &&
        +\frac{737 x^4 -14102 x^3 - 28209 x^2 + 610 x - 508}{1296 (x-1)^4}
\label{c81eff}
\eeqn

\subsubsection{Results and Comparisons}
To check the different components of the SM part, we have also
introduced the $B_{ij}$ functions \cite{KN}. These can be very
useful if one wants to introduce the effects of New Physics
through the Wilson coefficients defined at the scale $M_W$.
Dismissing any right-handed light quark operator and assuming
purely real contributions, the New Physics contribution can be
written as
\beqn
C_{7,8}^{(0,1)}=x_{7,8} \; C_{7,8}^{(0,1), {\rm SM}}(M_W)
\eeqn
then the $B_{ij}$ are the coefficients of the different $x_i$
factors (linear and quadratic, and a $x_i$ independent term). In
other words, the contribution of the New Physics can be expressed
as
\beqn
B_{s\gamma}^{NLO}&=&B_{22}\;+\;B_{27}\; x_7\;+\; B_{28}\; x_8
\;+\;B_{77}\;
x_7^2+\;B_{88}\; x_8^2 \;+\;B_{78} \;x_7 x_8\nonumber \\
\eeqn

Note the assumption in \cite{KN} that the proportionality factor
is the same for the LO and NLO. In our case we define a larger set
of $B_{ij}$ by allowing

\beqn
C_{7,8}^{(0,1)}=x_{7,8}^{0,1}\; C_{7,8}^{(0,1,) {\rm SM}}
\eeqn

As a check on the SM results note that we  exactly recover the
results for the Wilson coefficients at $M_W$ for both $C_{7,8}$
and at both LO and NLO (this also agrees with \cite{CDGG-smh}.
 More
satisfying is that we recover all the results of \cite{KN}. Below,
see Table~\ref{Bij}, we show our results for the coefficients
$B_{ij}$.  Here we switch to the default values of \cite{KN}
($\sqrt z=m_c/m_b=0.29$, $m_b=4.80$\,GeV, $m_t=175$\,GeV ,
$\alpha_s(m_Z)=0.118$, $|V_{ts}^* V_{tb}|/|V_{cb}|=0.976$) and
$\mu_b =\bar\mu_b$

\begin{table}[thb]
\centerline{\parbox{14cm}{\caption{\label{Bij}\small\sl Values of
the coefficients $B_{ij}(\delta)$ in units of $10^{-4}$, for
different choices of $\mu_b$}}}
\begin{center}
\begin{tabular}{|cc|cccccc|c|}
\hline $\mu_b$ & $\delta$ & $B_{22}$ & $B_{77}$ & $B_{88}$ &
$B_{27}$
 & $B_{28}$ & $B_{78}$ & $\sum B_{ij}$ \\
\hline\hline $m_b/2$
 & 0.90 & 1.322 & 0.335 & 0.015 & 1.265 & 0.179 & 0.074 & 3.190 \\
 & 0.30 & 1.169 & 0.322 & 0.005 & 1.196 & 0.136 & 0.070 & 2.898 \\
 & 0.15 & 1.081 & 0.309 & 0.004 & 1.144 & 0.126 & 0.067 & 2.730 \\
\hline $m_b$
 & 0.90 & 1.258 & 0.382 & 0.015 & 1.395 & 0.161 & 0.083 & 3.293 \\
 & 0.30 & 1.239 & 0.361 & 0.005 & 1.387 & 0.137 & 0.080 & 3.210 \\
 & 0.15 & 1.200 & 0.347 & 0.004 & 1.354 & 0.132 & 0.077 & 3.114 \\
\hline $2 m_b$
 & 0.90 & 1.023 & 0.428 & 0.015 & 1.517 & 0.132 & 0.092 & 3.206 \\
 & 0.30 & 1.041 & 0.402 & 0.004 & 1.552 & 0.118 & 0.091 & 3.209 \\
 & 0.15 & 1.021 & 0.386 & 0.004 & 1.534 & 0.115 & 0.088 & 3.149 \\
\hline
\end{tabular}
\end{center}
\end{table}


Switching back to our default central values, with $m_b=4.8$GeV,
$|V_{ts}^* V_{tb}|/|V_{cb}|^2=0.971$, $m_c=1.25$GeV, $m_t=174.3$,
$\alpha_s(M_Z^2)=0.1185$ and with $\mu_b=\bar
\mu_b=m_b=m_b^{1s}=4.80$GeV we find

\beqn
10^{4}\;B_{s\gamma}^{NLO}(\delta=0.9,\sqrt{z}=0.22)&=&1.512\;+\;1.417\;
x_7^0\;+\;0.155\; x_8^0 \nonumber \\
&+&0.136\; x_7^1\;+\;0.017 \;x_8^1\;+\;0.283\;
(x_7^0)^2+\;0.014\; (x_8^0)^2 \;+\;0.064 \;x_7^0 x_8^0\nonumber \\
&+& 0.103\; x_7^0 x_7^1 +0.013 \; x_7^0 x_8^1 \; \;+\;0.007 \;
x_8^0 x_7^1\;+\;0.001\; x_8^0 x_8^1
\eeqn
whereas in the assumption of \cite{KN} we get
\beqn
10^{4}\;B_{s\gamma}^{NLO}(\delta=0.9,\sqrt{z}=0.22)&=&1.512\;+\;1.553\;
x_7\;+\;0.173\; x_8 \;+\;0.386\;
x_7^2 \nonumber \\
&+&\;0.015\; x_8^2 \;+\;0.084 \;x_7 x_8
\eeqn
leading to
\beqn
B_{s\gamma}^{NLO,SM}(\delta=0.9,\sqrt{z}=0.22)=3.723 \; 10^{-4}
\eeqn
This result  agrees at the $2$ per-mil with the more sophisticated
analysis of \cite{Gambino-Misiak} and is in very good agreement
with the current experimental value.

\subsection{Charged Higgs contribution}
The charged Higgs contribution in our code is estimated at the
$M_W$ scale and is based on \cite{CDGG-smh}, therefore we neglect
any small running from $M_{H^\pm}$ to $M_W$. Indeed either
$M_{H^\pm}$ is very large in which case the Higgs contribution is
too small, or there is little running in a region where $\alpha_s$
is not too large.

\subsubsection{LO}
We have 
\beq
C^{(0) H^\pm}_{7,8}(\mu_W)=F_{7,8}^{2}(x_{Ht})\;+\;
\frac{1}{\tb^2} \frac{1}{3}F_{7,8}^{1}(x_{Ht})
\eeq
with
\beq
x_{Ht}=\frac{\overline{m}_t^2(\mu_W )}{M_H^2}~,
\eeq
and
\beqn
F_7^{(2)}(x)=\frac{x(3-5x)}{12(x-1)^2}+\frac{x(3x-2)}{6(x-1)^3}\ln
x \nonumber \\
F_8^{(2)}(x)=\frac{x(3-x)}{4(x-1)^2}-\frac{x}{2(x-1)^3}\ln x
\eeqn

\subsubsection{NLO}
We have
\beqn
 C_7^{(1)H^\pm}(\mu_W) &=& G_7^H(x_{Ht}) +
       \Delta_7^H(x_{Ht}) \ln\frac{\mu_W^2}{M_H^2}-\frac49 E^H(x_{Ht}) \nonumber\\
 C_8^{(1)H^\pm}(\mu_W) &=& G_8^H(x_{Ht}) +
       \Delta_8^H(x_{Ht}) \ln\frac{\mu_W^2}{M_H^2} -\frac16 E^H(x_{Ht})
\eeqn

\beqn
G_7^H(x) &= & -\frac{4}{3}x\left[
\frac{4(-3+7x-2x^2)}{3(x-1)^3}{\rm Li}_2 \left( 1 - \frac{1}{x}
\right)+ \frac{8-14x-3x^2}{3(x-1)^4}\ln^2x \right. \nonumber \\ &&
\left. +\frac{2(-3-x+12x^2-2x^3)}{3(x-1)^4}\ln x
+\frac{7-13x+2x^2}{(x-1)^3}\right] \nonumber \\ &&
+\frac{1}{\tb^2}\frac{2}{9}x\left[
\frac{x(18-37x+8x^2)}{(x-1)^4}{\rm Li}_2 \left( 1 - \frac{1}{x}
\right)+ \frac{x(-14+23x+3x^2)}{(x-1)^5}\ln^2x \right. \nonumber
\\ &&
 +\frac{-50+251x-174x^2-192x^3+21x^4}{9(x-1)^5}\ln x
\nonumber \\ && \left .
+\frac{797-5436x+7569x^2-1202x^3}{108(x-1)^4}\right]
\label{higceq}
\eeqn
\beqn
\Delta_7^H(x) &= & -\frac{2}{9}x\left[ \frac{21-47x+8x^2}{
(x-1)^3}+\frac{2(-8+14x+3x^2)}{(x-1)^4}\ln x \right] \nonumber \\
&& +\frac{1}{\tb^2}\frac{2}{9}x\left[ \frac{-31-18x+135x^2-14x^3}{
6(x-1)^4}+\frac{x(14-23x-3x^2)}{(x-1)^5}\ln x \right]
\eeqn
\beqn
G_8^H(x) &= & -\frac{1}{3}x\left[
\frac{-36+25x-17x^2)}{2(x-1)^3}{\rm Li}_2 \left( 1 - \frac{1}{x}
\right)+ \frac{19+17x}{(x-1)^4}\ln^2x \right. \nonumber \\ &&
\left. +\frac{-3-187x+12x^2-14x^3)}{4(x-1)^4}\ln x
+\frac{3(143-44x+29x^2)}{8(x-1)^3}\right] \nonumber \\ &&
+\frac{1}{\tb^2}\frac{1}{6}x\left[
\frac{x(30-17x+13x^2)}{(x-1)^4}{\rm Li}_2 \left( 1 - \frac{1}{x}
\right)- \frac{x(31+17x)}{(x-1)^5}\ln^2x \right. \nonumber \\ &&
 +\frac{-226+817x+1353x^2+318x^3+42x^4}{36(x-1)^5}\ln x
\nonumber \\ && \left.
+\frac{1130-18153x+7650x^2-4451x^3}{216(x-1)^4}\right]
\eeqn
\beqn
\Delta_8^H(x) &= &  -\frac{1}{3}x\left[ \frac{81-16x+7x^2}{
2(x-1)^3}-\frac{19+17x}{(x-1)^4}\ln x \right] \nonumber \\ &&
+\frac{1}{\tb^2}\frac{1}{6}x\left[ \frac{-38-261x+18x^2-7x^3}{
6(x-1)^4}+\frac{x(31+17x)}{(x-1)^5}\ln x \right]
\eeqn
\beq
E^H(x)=\frac{1}{\tb^2}\left[
\frac{x(16-29x+7x^2)}{36(x-1)^3}+\frac{x(3x-2)}{6(x-1)^4} \ln x
\right] ~.
\eeq
As a check we recover exactly the LO and NLO values quoted in
Table~1 of \cite{CDGG-smh}.

\subsection{SUSY contributions}
We only consider the contribution from charginos (and accompanying
squarks). Here we follow \cite{DGG} rather closely but  adapt the
expressions for the $\eps_b,\eps_b(t)$.

\subsubsection{LO}
We first consider the LO SUSY contribution at the SUSY scale.
Though the code allows to choose this scale, our default value is
set at the gluino mass $\mu_{\rm susy}=m_{\tilde{g}}$. This is
also the scale we take for the SUSY $\Delta m_b$ corrections.

With
\beqn
F_7^{(3)}(x)&=&\frac{5-7x}{6(x-1)^2}+\frac{x(3x-2)}{3(x-1)^3}\ln x
,
\label{f73}\\
F_8^{(3)}(x)&=&\frac{1+x}{2(x-1)^2}-\frac{x}{(x-1)^3}\ln x~.
\label{f83}
\eeqn

\bea
C_{7,8}^\chi(\mu_{SUSY})&=&
 \sum_{a=1,2}\left\{ \frac{2}{3} \frac{M_W^2}{\msq^2}
\tilde{V}_{a1}^2  F^{(1)}_{7,8}(x_{\tilde{q}\, \chi^+_a}) \right. \nonumber \\
&-& \frac2{3}  \left( \ct\,\tilde{V}_{a1}  - \st\, \tilde{V}_{a2}
\frac{m_t}{\sqrt{2}\sin\beta\,\mw} \right)^2
\frac{\mw^2}{\mst{1}^2} F_{7,8}^{(1)}(x_{{\tilde{t}_1}\,\chi^+_a}) \nonumber \\
&-& \frac2{3}  \left( \st\,\tilde{V}_{a1}  + \ct\, \tilde{V}_{a2}
\frac{m_t}{\sqrt{2}\sin\beta\,\mw} \right)^2
\frac{\mw^2}{\mst{2}^2} F_{7,8}^{(1)}(x_{{\tilde{t}_2}\,\chi^+_a})\nonumber \\
&+&  \frac{1}{\cos\beta} \left(
\frac{\tilde{U}_{a2}\tilde{V}_{a1}\mw}{\sqrt{2} m_{\chi^+_a}}
\left[F^{(3)}_{7,8}(x_{\tilde{q}\, \chi^+_a})- \ct^2 \,
F^{(3)}_{7,8} (x_{{\tilde{t}_1}\,\chi^+_a})
-\st^2 \,F^{(3)}_{7,8} (x_{{\tilde{t}_2}\,\chi^+_a}) \right] \right. \nonumber \\
&+&\left.  \left. \st \, \ct \frac{\tilde{U}_{a2}\, \tilde{V}_{a2}
\,m_t}{2\sin\beta \,m_{\chi^+_a}} \left[ F^{(3)}_{7,8}
(x_{{\tilde{t}_1}\,\chi^+_a}) - F^{(3)}_{7,8}
(x_{{\tilde{t}_2}\,\chi^+_a})\right]
 \right) \right\} ~~.
\eeqn
with obvious notations for the sparticles. $x_{ij}=m_i^2/m_j^2$
here and in the following. Our diagonalising matrices for the
chargino is as in \cite{DGG} ( as well as our convention for the
sign of $\mu$)
\beq
\tilde{U} \pmatrix{ M_2 & \mw \sqrt{2} \sin \beta \cr
              \mw \sqrt{2} \cos \beta & \mu } \tilde{V}^{-1}\,
\eeq

The squark mixing and definitions are also as in \cite{DGG}
\beqn
c_{\tilde{q}} &\equiv& \cos \theta_{\tilde{q}} \;\;,\;\;
s_{\tilde{q}} \equiv \sin\theta_{\tilde{q}} \;
\; {\rm with} \nonumber \\
{\tilde q}_1&=&c_{\tilde{q}} \,{\tilde q}_L + s_{\tilde{q}}
\,{\tilde q}_R \;,\; {\tilde q}_2=- s_{\tilde{q}} \, {\tilde q}_L
+c_{\tilde{q}} \,{\tilde q}_R \;,\;{\rm and}\;\; m_{{\tilde q}_1}>
m_{{\tilde q}_2}
\eeqn

\subsubsection{$\Delta m_b$ corrections and large $\tb$ effects}
\beq
m_b=\sqrt{2}M_W~\frac{y_b}{g}~\cos\beta \left( 1+\epsilon_b
\tan\beta \right)\;\;,\;\; \delta m_b=\frac{\Delta
m_b}{m_b}=\epsilon_b \tb . \label{massab}
\eeq

We implement $\delta m_b$ in our code as follows. First define
\beq
H_2(x,y)  = \frac{x\ln\, x}{(1-x)(x-y)} + \frac{y\ln\, y}{(1-y)
(y-x)}~\;\;\;\; H(i,j,k)=H_2(x_{ik},y_{jk})
\eeq

where $(x,y)_{ij}=m_i^2/m_j^2$.

\beqn
B(m_1,m_2,Q^2)=\frac{1}{2}\left(
\frac{1}{2}+\frac{1}{1-x}+\frac{\ln
x}{(1-x)^2}-\ln(m_2^2/Q^2)\right) \;\;,\;\;x=m_2^2/m_1^2
\eeqn

Then

\beqn
\delta m_b&=&\frac{2\,\alpha_s(\mu_{susy})}{3\,\pi} \left(
\frac{A_b-\mu \tb}{\mg} H(\tilde{b}_1,\tilde{b}_2,\tilde{g})
-\frac{1}{2}(B(\tilde{g},\tilde{b}_1,\mu_{susy}^2)+B(\tilde{g},\tilde{b}_2,\mu_{susy}^2))
\right)
\nonumber \\
&+&
           \frac{ \tilde{y}_t^2(\mu_{susy})}{16\, \pi^2} \,\sum_{a=1,2} \tilde{U}_{a 2}\frac{\mu-A_t \tb}{m_{\chi^+_a}}
\,H(\tilde{t}_1,\tilde{t}_2,\chi^+_a) \,
\tilde{V}_{a 2} \nonumber \\
&+& \frac{\alpha(M_Z)}{4 s_W^2 \pi} \mu M_2 \tb
\left(\frac{\ct^2}{\msto^2}
H(M_2,\mu,\tilde{t}_1)+\frac{\st^2}{\mstt^2}
H(M_2,\mu,\tilde{t}_2) \right. \nonumber
\\ & & \;\;\;\;\;\hspace*{3cm} + \left. \frac{c_{\sbottom}^2}{2
\msbo^2} H(M_2,\mu,\tilde{b}_1)+\frac{s_{\sbottom}^2}{2 \msbt^2}
H(M_2,\mu,\tilde{b}_2)\right) \label{epsb}
\eeqn

Note that we have included the electroweak contributions with
enhanced $\tb$. For this part we make the approximation that the
masses of the charginos are given by $M_2$ and $\mu$, and neglect
the mixing matrices, we have also neglected the $U(1)$
contribution which is even smaller. Although the formulae above
include non-leading $\tb$ effects, for consistency we have not
coded these contributions in our program. The electroweak
corrections agree with those in \cite{Carena} whereas  the sbottom
term is missing in \cite{Pierce}. However we find the $SU(2)$
gauge contribution to be rather small compared to the strong and
Yukawa contributions.

We turn now to $\epsilon_b^\prime (t)$
\beqn
\epsilon_b^\prime (t)&=& +\frac{2\,\alpha_s(\mu_{susy})}{3\,\pi}
\frac{A_b/\tb-\mu}{\mg}
   \left[ \ct^2 c^2_{\tilde{b}}
\,H(\tilde{t}_1,\tilde{b}_2, \tilde{g}) + \ct^2 s^2_{\tilde{b}}
\,H(\tilde{t}_1, \tilde{b}_1, \tilde{g}) + \right.
\nonumber \\
&& \left. ~~~~~~~~~~~~~~~ \st^2 c^2_{\tilde{b}} \,
H(\tilde{t}_2,\tilde{b}_2, \tilde{g}) + \st^2 s^2_{\tilde{b}} \,
H(\tilde{t}_2,\tilde{b}_1, \tilde{g})
\right]  \nonumber \\
&& + \frac{ y_t^2(\mu_{susy})}{16\, \pi^2} \sum_{a=1}^{4}\,N^*_{a
4 }\frac{A_t-\mu/\tb}{m_{\chi^0_a}}
   \, \left[ \ct^2 c^2_{\tilde{b}}\,
    H(\tilde{t}_2,\tilde{b}_1,\chi^0_a) +
    \ct^2 s^2_{\tilde{b}}\,
    H(\tilde{t}_2,\tilde{b}_2,\chi^0_a) + \right.
\nonumber \\
&&\left.~~~~~~~~~~~~~~~\st^2 c^2_{\tilde{b}}\,
    H(\tilde{t}_1,\tilde{b}_1,\chi^0_a) +
    \st^2 s^2_{\tilde{b}}\,
    H(\tilde{t}_1,\tilde{b}_2,\chi^0_a) \right]
             \, N_{a 3}  \nonumber \\
&+& \frac{\alpha(M_Z)}{4 s_W^2 \pi} \mu M_2  \left(
\frac{c_{\sbottom}^2}{\msbo^2}
H(M_2,\mu,\tilde{b}_1)+\frac{s_{\sbottom}^2}{\msbt^2}
H(M_2,\mu,\tilde{b}_2) \right. \nonumber
\\ & & \;\;\;\;\;\hspace*{3cm} + \left.
\frac{\ct^2}{2 \msto^2} H(M_2,\mu,\tilde{t}_1)+\frac{\st^2}{2
\mstt^2} H(M_2,\mu,\tilde{t}_2) \right) \label{epsbt}
\eeqn

Note that we have added an electroweak contribution, in the same
approximation as in $\delta m_b$. Most importantly, the sign of
the Yukawa contribution and the elements of the  diagonalising
matrices that appear in Eq.~\ref{epsbt} are
 different from those in \cite{DGG}. We have verified this
by explicit calculation of $\epsilon_b^\prime (t)$, moreover with
our formula, we find that in the decoupling limit we do indeed
have $\epsilon_b^\prime (t) \ra \epsilon_b$, which would not have
been the case had we blindly used the formula of \cite{DGG}. One
of the authors of \cite{DGG}, Paolo Gambino, has recently
confirmed our implementation.

As for $\epsilon_t^\prime (s)$, we find that only the QCD
contribution remains. Note, as said previously, we shall only keep
the $\tb$ enhanced term in our code.

For $\epsilon_t^\prime (s)$
\beqn
\epsilon_t^\prime (s)&=& -\frac{2\,\as}{3\,\pi}
\frac{\mu+A_t/\tb}{\mg} \left[ \ct^2
\,H(\tilde{t}_2,\tilde{s},\tilde{g}) \;+\; \st^2
H(\tilde{t}_1,\tilde{s},\tilde{g}) \right]
\eeqn

We find no $\tb$ enhanced electroweak gauge contribution.

Once more let us stress that, although we have derived, for the
$\eps, \eps^\prime$ the $\tb$ enhanced and the non $\tb$ enhanced
we will only keep the $\tb$ enhanced terms, since these are the
ones that can be resummed. Therefore in our numerical analysis
that takes this resummation into account we only keep the $\tb$
enhanced terms.

As we mentioned earlier these $\eps$'s contributions are to be
evaluated at the scale $Q^2=\mu_{\rm susy}^2>m_t^2$ which we
associate with the gluino mass. In particular $\alpha_s$ is to be
evaluated here taking into account 6 active quarks.
\beqn
\eta_s \equiv \alpha_s(\mu_{SUSY})/\alpha_s(\mu_W)= \left(1
+\frac{7 \alpha_s(\mu_W)}{2 \pi} \ln(\mu_{SUSY}/
\mu_W)\right)^{-1}
\eeqn

Also the Yukawa (and top mass) that is used for the chargino
contribution is
\beqn
\tilde{y}_t (\mu_{SUSY}) &=& y_t (\mu_W) \left[
\frac{\alpha_s(\mu_{SUSY})}{\alpha_s(\mt)} \right]^{4/7}
\left[ \frac{\alpha_s(\mt)}{\alpha_s(\mu_W)} \right]^{12/23}\nonumber \\
& \times & \frac{1}{\sqrt{1+\frac{9y_t^2(m_t)}{8\pi\alpha_s(m_t)}
\left\{ \left[ \frac{\alpha_s(\mu_{SUSY})}{\alpha_s(\mt)}
\right]^{1/7}-1 \right\} }} \label{yina}
\eeqn

At all scales we relate $y_t$ to $m_t$ as follows
\beqn
y_t^2(Q^2)=\frac{2 \pi \alpha(M_Z)}{s_W^2} \frac{1+\tb^2}{\tb^2}
\frac{m_t^2(Q^2)}{M_W^2}
\eeqn
We used a fixed $\tb$ at all scales, we neglect the small running
of $\alpha$ between $M_Z$ and $\mu_{susy}$.

 Though the $\eps$ effects are to be extracted at
$\mu_{{\rm susy}}$, they are included in the SM and charged Higgs
Wilson coefficient at $\mu_W$
\beqn
\delta C_{7,8}^{(SM)}({\rm leading} \tan\beta)(\mu_W)&=&
\frac{\left[ \epsilon_b -\epsilon^\prime_b(t)\right] \tan\beta}{1
+\epsilon_b\tan\beta}
 \, F_{7,8}^{(2)}(x_{tw})
\label{tbSM} \\
\delta C_{7,8}^{(H^\pm)}({\rm leading} \tan\beta)(\mu_W)&=&
-\frac{\left[ \epsilon^\prime_t(s) +\epsilon_b\right] \tan\beta}{1
+\epsilon_b\tan\beta}
 \, F_{7,8}^{(2)}(x_{Ht}).
\label{tb2H}
\eeqn

For the chargino contribution we first evaluate the Wilson
coefficients at the scale $\mu_{\rm SUSY}$ by including the top
mass effect and the $\eps$ effects. In our implementation we
assume, as occurs in most cases in mSUGRA, that both stops are
heavy, so that they are decoupled together at the same scale
$\mu_{\rm SUSY}$,
\beqn
C_{7,8}^\chi(\mu_{SUSY})&=&
 \sum_{a=1,2}\left\{ \frac2{3} \frac{\mw^2}{\msq^2}
\tilde{V}_{a1}^2  F^{(1)}_{7,8}(x_{\tilde{q}\, \chi^+_a}) \right. \nonumber \\
&-& \frac2{3}  \left( \ct\,\tilde{V}_{a1}  - \st\, \tilde{V}_{a2}
\frac{\overline{m}_t(\mu_{SUSY})}{\sqrt{2}\sin\beta\,\mw}
\right)^2
\frac{\mw^2}{\mst{1}^2} F_{7,8}^{(1)}(x_{{\tilde{t}_1}\,\chi^+_a}) \nonumber \\
&-& \frac2{3}  \left( \st\,\tilde{V}_{a1}  + \ct\, \tilde{V}_{a2}
\frac{\overline{m}_t(\mu_{SUSY})}{\sqrt{2}\sin\beta\,\mw}
\right)^2
\frac{\mw^2}{\mst{2}^2} F_{7,8}^{(1)}(x_{{\tilde{t}_2}\,\chi^+_a})\nonumber \\
&+&  \frac{K_b}{\cos\beta} \left(
\frac{\tilde{U}_{a2}\tilde{V}_{a1}\mw}{\sqrt{2} m_{\chi^+_a}}
\left[F^{(3)}_{7,8}(x_{\tilde{q}\, \chi^+_a})- \ct^2 \,
F^{(3)}_{7,8} (x_{{\tilde{t}_1}\,\chi^+_a})
-\st^2 \,F^{(3)}_{7,8} (x_{{\tilde{t}_2}\,\chi^+_a}) \right] \right. \nonumber \\
&+&\left.  \left. \st \, \ct \frac{\tilde{U}_{a2}\, \tilde{V}_{a2}
\,\overline{m}_t(\mu_{SUSY})}{2\sin\beta \,m_{\chi^+_a}} \left[
F^{(3)}_{7,8} (x_{{\tilde{t}_1}\,\chi^+_a}) - F^{(3)}_{7,8}
(x_{{\tilde{t}_2}\,\chi^+_a})\right]
 \right) \right\} ~~.
\label{final}
\eeqn

with
\beqn
K_b = 1/(1 + \epsilon_b \tb)
\eeqn
These are then evolved to $\mu_W$ as
\beqn
C_7^\chi(\mu_W)&=& \eta^{-\frac{16}{3 \beta_0^\prime}}
C_7^\chi(\mu_{SUSY}) +\frac83 \left(
  \eta^{-\frac{14}{3 \beta_0^\prime}}-\eta^{-\frac{16}{3 \beta_0^\prime}}\right)
C_8^\chi(\mu_{SUSY})
\nonumber \\
C_8^\chi(\mu_W)&=& \eta^{-\frac{14}{3 \beta_0^\prime}}
C_8^\chi(\mu_{SUSY}) \label{c7rr}\;\;\; ;\;\;\; \beta_0^\prime=-7
\eeqn
At $\mu_W$ all the contributions ($SM$, $H^\pm$ including $\tb$
effects) are added together with those of the chargino
contribution and evolved according to Eq.~\ref{evolvemwmb}.

To check on the SUSY part and the implementation of $\tb$ enhanced
terms, we have first verified that we had perfect agreement with
Fig.~4 of \cite{DGG}. This  in fact is only a check on the
implementation of the $\eps$'s in the SM and $H^+$ contribution
with fixed $\eps$. A full check of the SUSY contribution requires
a quite large set of inputs. In \cite{DGG} one can read the effect
on $B\ra X_s \gamma$ of a SUGRA model. Since the outputs of SUGRA
codes can differ quite a bit, we have not used our own RGE but
requested the weak scale parameters from Paolo Gambino used for
their Fig.~2 and Fig.~3 in \cite{DGG} , which in passing includes
only the $\alpha_s$ contribution to the $\eps$. We have found an
excellent agreement both  for positive and negative $\mu$.

\end{document}